\documentstyle[eqsecnum,aps,pre,emlines2,bezier,epsfig]{revtex}

\begin{document}


\def\eps{\xi}
\def\const{{\rm const}}
\def\Dm{\widetilde{\cal D}_{\mu}}
\def\D{{\cal D}}
\def\B{\bbox{B}}
\def\BO{\bbox{B}^o}
\def\BP{\bbox{B}'}
\def\k{{\bf k}}
\def\q{{\bf q}}
\def\p{{\bf p}}
\def\bv{\bbox{v}}
\def\bfr{\bbox{r}}
\def\bfx{\bbox{x}}
\def\partt{\bbox{\partial}}


\def\gdot{\circle*{20}}
\def\dhline#1#2#3#4#5#6{  {\countdef\nnn=255
\dimendef\llx=0   \dimendef\lly=1
\dimendef\dx=2   \dimendef\dy=3
\llx=#1\unitlength  \lly=#2\unitlength
\dx=#4\unitlength   \dy=#5\unitlength  \nnn=#6
\divide\nnn by 2
\advance\dx by-\llx \advance\dy by-\lly
\div\nnn \div4 \lline \adv
\multiply\dx by2 \multiply\dy by2
\loop \adv \ifnum\nnn>1 \lline \adv \advance\nnn by-1
\repeat \div2 \lline }}

\def\div#1{ \divide\dx by#1  \divide\dy by#1 }
\def\adv{ \advance\llx by\dx \advance\lly by\dy }

\def\lline{ {  \divide\llx by\unitlength \divide\lly by\unitlength
\divide\dx by\unitlength \divide\dy by\unitlength
\advance\dx by\llx \advance\dy by\lly
\emline{\number\llx}{\number\lly}{}{\number\dx}{\number\dy}{}}}

\unitlength=0.04ex
\special{em:linewidth 0.3pt}

\def\dA{
\unitlength=0.04ex
\special{em:linewidth 0.3pt}
\begin{picture}(180,140)
\emline{10}{10}{}{90}{130}{}
\emline{90}{130}{}{170}{10}{}
\dhline{20}{25}{}{160}{25}{10}
\emline{38}{39}{}{26}{47}{}
\emline{154}{47}{}{142}{39}{}
\put(90,130){\gdot}
\end{picture}}

\def\dS{
\unitlength=0.04ex
\special{em:linewidth 0.3pt}
\begin{picture}(180,40)
\emline{10}{70}{}{170}{70}{}
\emline{23}{63}{}{23}{77}{}
\emline{130}{63}{}{130}{77}{}
\dhline{36}{70}{}{45}{82}{2}
\dhline{45}{82}{}{57}{92}{2}
\dhline{57}{92}{}{70}{99}{2}
\dhline{70}{99}{}{85}{103}{2}
\dhline{85}{103}{}{101}{103}{2}
\dhline{101}{103}{}{116}{99}{2}
\dhline{116}{99}{}{129}{92}{2}
\dhline{129}{92}{}{141}{82}{2}
\dhline{141}{82}{}{150}{70}{2}
\end{picture}}

\def\dSS{
\unitlength=0.04ex
\special{em:linewidth 0.3pt}
\begin{picture}(180,40)
\emline{10}{70}{}{170}{70}{}
\emline{23}{63}{}{23}{77}{}
\emline{160}{63}{}{160}{77}{}
\dhline{36}{70}{}{45}{82}{2}
\dhline{45}{82}{}{57}{92}{2}
\dhline{57}{92}{}{70}{99}{2}
\dhline{70}{99}{}{85}{103}{2}
\dhline{85}{103}{}{101}{103}{2}
\dhline{101}{103}{}{116}{99}{2}
\dhline{116}{99}{}{129}{92}{2}
\dhline{129}{92}{}{141}{82}{2}
\dhline{141}{82}{}{150}{70}{2}
\end{picture}}


\title{Persistence of small-scale anisotropies and anomalous scaling \\
in a model of magnetohydrodynamics turbulence}

\author{N.~V.~Antonov$^{1}$, A.~Lanotte$^{2}$, and A.~Mazzino$^{3,4}$\\
\small{$^{1}$ Department of Theoretical Physics, St Petersburg University,
Uljanovskaja 1,\\ St Petersburg, Petrodvorez, 198904 Russia\\
\small$^{2}$ CNRS, Observatoire de la C\^{o}te d$'$Azur, B.P. 4229,
06304 Nice Cedex 4, France\\
\small$^3$ INFM-Department of Physics, Genova University, Via Dodecaneso
33,
I--16142  Genova, Italy}\\
\small$^4$ The Niels Bohr Institute, Blegdamsvej 17, DK-2100 Copenhagen,
Denmark}

\draft
\maketitle

\begin{abstract}
The problem of anomalous scaling in magnetohydrodynamics
turbulence is considered within the framework of the kinematic
approximation, in the presence of a large-scale
background magnetic field.
The velocity field is Gaussian, $\delta$-correlated in time, and scales
with a positive exponent $\xi$.
Explicit inertial-range expressions for the magnetic correlation
functions are obtained; they are represented by superpositions of
power laws with non-universal amplitudes and universal (independent
of the anisotropy and forcing) anomalous exponents.
The complete set of anomalous exponents for the pair correlation
function is found non-perturbatively, in any space dimension $d$,
using the zero-mode technique.
For higher-order correlation functions, the anomalous exponents
are calculated to $O(\xi)$ using the renormalization group.
The exponents exhibit a hierarchy related to
the degree of anisotropy; the leading contributions to the even
correlation functions are given by the exponents from the isotropic shell,
in agreement with the idea of restored small-scale isotropy.
Conversely, the small-scale anisotropy reveals itself in the odd
correlation functions\,: the skewness factor is slowly decreasing going
down to small scales and higher odd dimensionless ratios
(hyperskewness etc.) dramatically increase, thus diverging in the
$r\to 0$ limit.
\end{abstract}

 \section{Introduction} \label{sec:Intro}

In cosmical objects, small-scale evolution of the magnetic field
$\bbox{B}$ often takes place in the presence of a strong large-scale
magnetic field $\bbox{B}^o$. It is, for example, what happens in the
solar corona where, in spite of the typical value of the sun magnetic field
($\approx 1$ Gauss), fields as intense as $\approx 500$ Gauss
can be observed in solar flares.
These highly energetic and large-scale events coexist with small-scale
turbulent activity, finally responsible of the dissipation of magnetic
field
energy. Modelling the way through which energy is stored and then
dissipated is, consequently, not an easy task.

In Ref.~\cite{EVPP96}, the following description is proposed: a
large-scale axial, e.g., directed parallel to some vector $\hat{\bbox
{z}}$,
magnetic field $\bbox{B}^o$ is assumed to dominate the dynamics in
the $\hat{\bbox{z}}$ direction, while
the activity in the transverse plane can be satisfactorily described
as quasi-bidimensional. This picture allows reliable numerical
simulations in two dimensions, from
which it appears clear that the magnetic field tends to organize
in rare large-scale structures separated by narrow current sheets.
Deep investigation of small-scale intermittency properties is still
not permitted by lack of spatial resolution.

An interesting question raised by this problem, beside structure
formation, is related to the role played by large-scale anisotropy
on the small-scale statistics.
Indeed, this is quite a typical situation in turbulence, where almost
every large-scale forcing is not isotropic. Here, instead of taking
the restoration of local small-scale isotropy for granted, as in the
Kolmogorov theory of turbulence \cite{K41,Monin,F95},
we analyze in detail the effects of anisotropic large-scale
contributions on the small-scale magnetic fluctuations.

A wide interest has been recently devoted to this issue
\cite{Sree,synth,Sadd94,PS95,Siggia,Pumir,BoOr96,Arad98,Arad99,%
A98,LM99,CLMV99}.
 From the viewpoints of theoretical and numerical analysis,
focusing on a small number of indicators
some arguments are given in favor of the small-scale isotropy
restoration  in the Navier--Stokes (NS) turbulence \cite{BoOr96,Arad99}.
On the other hand, investigating a larger class of anisotropic indicators,
footprints of small-scale anisotropy become manifest \cite{PS95}.
The scenario thus appears extremely faceted and needs of further
investigations.

Recently, clear evidences of persistent small scale
anisotropy have been found in Ref.~\cite{CLMV99}, where the statistical
properties of a scalar field advected by the non-intermittent NS
flow generated in a two-dimensional inverse cascade regime are
investigated.

Two main goals motivate this paper. On one hand, we give
details of the results presented in the Rapid Communication \cite{LM99}
where
the effects of anisotropy on scaling exponents of the two-point
magnetic field correlations have been addressed in the framework
of the kinematic magnetohydrodynamics (MHD) problem. Non-perturbative
expressions for the scaling exponents were derived and their universality
proved. Specifically, there arises a picture
of a non trivial statistical behavior, where anisotropic fluctuations
are organized in a hierarchical order according to their degree of
anisotropy.
Contributions belonging to shells of higher anisotropic index
decay faster, and the isotropic contribution finally dominates.

However, the dominance of the isotropic contribution in the scaling
exponents
does not imply that large-scale anisotropy is irrelevant for the
small-scale magnetic statistics. A deep investigation focused on a larger
number of statistical indices (that is focused on the proper anisotropy
indicators) has to be performed in order to highlight the way (if any)
through which large-scale anisotropy manifests itself at small scales.
This is the second aim of the present paper.
Specifically, in addition to the non-perturbative results for
the two-point correlations, we present new results dealing
with higher-order magnetic correlation functions. Being more specific,
we exploit the field theoretic renormalization group (RG) to obtain
the anomalous exponents for higher-order magnetic correlation functions
at the first order in $\xi$, the exponent entering into
the velocity covariance. In particular, we evaluate
the odd-order correlation function exponents, from which dimensionless
ratios like skewness and hyper-skewness are calculated.
As a result, in three dimensions, the former behaves at the
dissipative scale as $Pe^{-1/10}$ while the latter as
$Pe^{11/10}$, $Pe$ being the P\'eclet number (i.e.,~the equivalent of the
Reynolds number for the NS turbulence).
Notice the opposite signs appearing in the scaling exponents. They are
the signature
of persistent small-scale activities. Indeed, the first index is weakly
scale dependent while the second is even divergent at small scales
(i.e.,~$Pe\to\infty$). Let us remark that to restore isotropy at small
scales
all such indices should decay to zero as $Pe$ grows.

The same general picture is found numerically in Ref.~\cite{CLMV99}
in the framework of the passive scalar advection by NS flows.
In addition, the results are in qualitative agreement with the first-order
analytic expressions for the anomalous exponents obtained in \cite{A98}
for the passive scalar advected by a synthetic velocity field.

The paper is organized as follows.
In Sec.~\ref{sec:Def}, we give the detailed definition of the
kinematic MHD Kasantzev--Kraichnan model, which describes the
passive advection of the magnetic field by the Gaussian white-in-time,
self-similar velocity field.
In Sec.~\ref{sec:FT}, the field theoretic formulation of the model is
presented. It allows for the derivation of the closed exact equations
for the response function and equal-time pair correlation
function
of the magnetic field. From the homogeneous solutions (zero modes)
of the pair correlation equation, scaling exponents of the pair correlation
function are determined.
In Sec.~\ref{sec:2-point}, these exponents are found
non-perturbatively, for any $\xi$ and space dimensionality $d$.
In Sec.~\ref{sec:RG}, we discuss the
UV renormalization of the model and derive the corresponding
beta functions and RG equations. The latter
possess an infrared (IR) stable fixed point,
which establishes the existence of anomalous scaling for
all the higher-order correlation functions. The inertial-range behavior
of these functions is determined by the scaling dimensions of
certain tensor composite operators; they are calculated in
Sec.~\ref{sec:Operators} to the first order in $\xi$
(one-loop approximation).
In Sec.~\ref{sec:OPE}, we employ the operator product expansion to
give explicit inertial range expressions for various
higher-order correlation functions. The results obtained are reviewed
in Sec.~\ref{sec:Conclusion}, where a brief comparison with
the passive scalar problem is also given.

 \section{Definition of the kinematic MHD Kasantzev--Kraichnan model}
 \label {sec:Def}

In the presence of a mean component $\bbox{B}^o$ (actually supposed to be
varying on a very large scale $\sim L$, the largest one in our problem)
the kinematic MHD equations describing the evolution of the fluctuating
part $\bbox{B} \equiv \bbox{B}(x)$
of the magnetic field are \cite{Zeldo83}:
\begin{eqnarray}
\label{fp}
\partial_t B_\alpha +\bbox{v}\cdot\bbox{\partial}\,
B_\alpha=\bbox{B}\cdot\bbox{\partial}\,v_{\alpha}+
\bbox{B}^o\cdot \bbox{\partial}\,v_{\alpha}+
\kappa_{0}\,\partial^2 B_{\alpha},\qquad
\alpha=1,\cdots ,d .
\end{eqnarray}
Here and below $x\equiv\{t,\bfx\}$,
$\partt\equiv\{\partial_{\alpha}=\partial/\partial x_{\alpha}\}$,
$\partial^2 \equiv \partial_{\alpha}\partial_{\alpha}$ is the
Laplace operator, $d$ is the dimensionality of the ${\bfx}$ space
and $\bbox{v}=\bbox{v}(x)$ is the velocity field. Both $\bbox{v}$
and $\bbox{B}$ are divergence free (solenoidal) vector fields:
$\partial_{\alpha} v _{\alpha} = \partial_{\alpha} B_{\alpha} =0$.\\
Equation (\ref{fp}) follows from the simplest form of Ohm's law for
conductive moving medium,
$\bbox{j} = \sigma\, (\bbox{E}+ \bbox{v}\times \bbox{B}/c)$, and the
Maxwell equations neglecting the displacement current:
$\partial_{t} \bbox{B} /c  + \partt\times \bbox{E} =0$,
$\partt\times \bbox{B} = 4\pi \bbox{j} /c =0$ and
$\bbox{\partial}\cdot \bbox{B}=0$. Here $c$ is the speed of light,
$\bbox{j}$ is the density of the electric current,
$\sigma$ is the conductivity, and $\kappa_{0}\equiv c^{2}/4\pi\sigma$
is the magnetic diffusivity.\\
The term $\bbox{B}^o\cdot\bbox{\partial}\,v_{\alpha}$ in (\ref{fp})
effectively
plays the same role as an external forcing driving the system and
being also a source of anisotropy for the magnetic field statistics.

In the real problem, $\bbox{v}$ obeys the NS equation with the
additional Lorentz force term
$\propto (\bbox{\partt}\times \bbox {B})\times \bbox {B}$, which
describes the effects of the magnetic field on the velocity field.
The framework of our analysis is the {\it kinematic}
MHD problem, where the reaction of the magnetic field
$\bbox{B}$ on the velocity field $\bbox{v}$ is neglected.
We assume that at the initial stages $\bbox{B}$ is weak and does not
affect the motions of the conducting fluid: it becomes then
a natural assumption to consider the dynamics
linear in the magnetic field strength \cite{Zeldo83}.
It is also noteworthy that in more realistic models of the MHD
turbulence the magnetic field
indeed behaves as a passive vector in the so-called kinetic fixed
point of the RG equations (see Refs.~\cite{MHD1,MHD2}).

For general velocity fields the well-known closure problem arises
even for the kinematic model.
This means that the equations of evolution for the single-time
multiple-space moments such as
$\left\langle B_{\alpha}(t,\bbox{r}_1)\cdots B_{\lambda}(t,
\bbox{r}_n)\right\rangle$ are not closed.
The situation changes for white-in-time random velocity fields.
The physical choice of a real turbulent flow governed by the NS
equation is then replaced by an incompressible, self-similar
advecting field, with Gaussian statistics and rapidly changing
($\delta$-correlated) in time. This last property
 allows us to write closed equations for the moments of
the magnetic field $\bbox{B}$ and to perform analytical
(both perturbative and non-perturbative) approaches to the $d$-dimensional
problem. Indeed, in the presence of a white-in-time random velocity field,
the solution is a Markov process in the time variable
and closed moment equations, sometimes called ``Hopf equations'',
can be obtained in analogy to the passive scalar case \cite{K68}.
Such models have attracted enormous attention recently
(see, e.g., Refs.~\cite{Pumir,K94,GK,Falk1,FMV98,RG} and references
therein) because of the insight they offer into the origin of
intermittency and anomalous scaling in fully developed turbulence.
We also note that the isotropic version of the kinematic rapid-change
magnetic model dates back to 1967 (see Ref.~\cite{KA68}) and was
studied by Refs. \cite{V96,RK97,AA98,CFKV}.

More precisely, we shall consider a simplified model in which ${\bv}(x)$
is a Gaussian random field, homogeneous, isotropic and
white-in-time, with zero mean and covariance
\begin{mathletters}
\label{2-point-v}
\begin{equation}
\langle v_{\alpha}(x) v_{\beta}(x')\rangle = \delta(t-t')\,
K_{\alpha\beta}(\bbox{r})
\label{temporal}
\end{equation}
with
\begin{equation}
K_{\alpha\beta} (\bbox{r}) = D_{0}\,
\int \frac{d{\bf k}}{(2\pi)^d} \,
\frac{P_{\alpha\beta}({\bf k})} {k^{d+\eps}} \,
\exp [{\rm i}{\bf k}\cdot\bbox{r}] ,
\quad \bbox{r}\equiv {\bfx}-{\bfx'},
\label{spatial}
\end{equation}
\end{mathletters}
where $P_{\alpha\beta}({\bf k})=\delta_{\alpha\beta}-k_\alpha k_\beta/
k^2$ is the transverse projector, $\k$ is the momentum, $k\equiv|{\bf k}|$,
$D_{0}>0$ is an amplitude factor, and $0<\eps<2$
is a free parameter. The IR regularization is provided by
the cut-off in the integral (\ref{2-point-v}) from below at
$k\simeq m$, where $m\equiv 1/L$ is the reciprocal of the
integral turbulence scale; the precise form of the cut-off
is not essential. For $0<\eps<2$, the difference
\begin{equation}
S_{\alpha\beta}(\bbox{r}) \equiv K_{\alpha\beta} ({\bf 0}) -
K_{\alpha\beta} (\bbox{r})
\label{difference}
\end{equation}
has a finite limit for $m\to0$:
\begin{equation}
S_{\alpha\beta}(\bbox{r})=
D r^{\xi}\left [\left (d+\xi-1\right)
\delta_{\alpha\beta} - \xi \frac{r_{\alpha}r_{\beta}}{r^2}
\right],
\label{eddydiff}
\end{equation}
with
\[ D = \frac{- D_{0}\, \Gamma(-\eps/2)} {(4\pi)^{d}2^{\eps}(d+\eps)
\Gamma(d/2+\eps/2)}, \]
where $\Gamma(\cdots)$ is the Euler gamma function
(note that $D>0$). It follows from Eq. (\ref{eddydiff})
that $\eps$ can be viewed as a kind of H\"{o}lder exponent,
which measures the {\em roughness\/} of the velocity field.
In the RG approach, the exponent $\eps$ plays the same
role as the parameter $\varepsilon=4-d$ does in the RG theory
of critical phenomena \cite{Zinn}. The relations
\begin{equation}
g_{0} \equiv D_{0}/\kappa_0 \equiv \Lambda^{\eps}
\label{g0}
\end{equation}
define the coupling constant $g_{0}$ (i.e., the expansion parameter
in the ordinary perturbation theory) and the characteristic
ultraviolet (UV) momentum scale $\Lambda$.

 \section{Field theoretic formulation of the model. Dyson equations for
 the pair correlation functions}
 \label {sec:FT}

The stochastic problem (\ref{fp}), (\ref{2-point-v})
is equivalent
to the field theoretic model of the set of three fields
$\Phi\equiv\{\BP, \B, {\bv}\}$ with action functional
\begin{equation}
S(\Phi)= \BP \Bigl[ - \partial_{t}\B -({\bv}\cdot\partt) \B +
(\B\cdot\partt){\bv}  + (\BO\cdot\partt){\bv}
+ \kappa _0\partial^{2} \B
\Bigr] -{\bv} K^{-1} {\bv}/2.
\label{action}
\end{equation}
The first five terms represent the Martin--Siggia--Rose action
(see, e.g., Refs.~\cite{Zinn,UFN,turbo})
for the stochastic problem (\ref{fp}) at fixed ${\bv}$,
and the last term represents the Gaussian averaging over ${\bv}$;
$K^{-1}$ is the inverse integral operation for (\ref{spatial}) and
$\BP$ is a solenoidal response vector field.
In (\ref{action}) and analogous formulas below, the required
integrations over $\{t,{\bfx}\}$ and summations over the vector
indices are implied, for example,
\[ \BP \partial_{t}\B \equiv \int\! dt\! \int\! d {\bfx}
\, B_{\alpha}' (x) \partial_{t} B_{\alpha}(x), \quad
{\bv} K^{-1} {\bv} \equiv \int \!dt\! \int\! d {\bfx}\! \int\!
d {\bfx}' \, v_{\alpha} (t,{\bfx} )  K_{\alpha\beta}^{-1}
({\bfx}-{\bfx}') v_{\beta} (t,{\bfx}'). \]

The formulation (\ref{action}) means that statistical averages of random
quantities in the stochastic problem (\ref{fp}), (\ref{2-point-v})
coincide with functional averages with the weight $\exp S(\Phi)$.
The model (\ref{action}) corresponds to a standard Feynman
diagrammatic technique with the triple vertex
$ \BP \left[-({\bv}\cdot\partt) \B + (\B\cdot\partt){\bv} \right] =
B'_{\alpha} B_{\beta} v_{\gamma} V_{\alpha\beta\gamma}$
with vertex factor
\begin{eqnarray}
V_{\alpha\beta\gamma}(\k,\p,\q)
= {\rm i} k_{\gamma} \delta_{\alpha\beta}
- {\rm i} k_{\beta} \delta_{\alpha\gamma} =
- {\rm i} p_{\gamma} \delta_{\alpha\beta}
+ {\rm i} q_{\beta} \delta_{\alpha\gamma},
\label{vertex}
\end{eqnarray}
where $\k$, $\p$ and $\q$ are the momenta flowing into the vertex via
the fields $\BP$, $\B$ and ${\bv}$, respectively.
Strictly speaking, the vertex $V_{\alpha\beta\gamma}$ has to be
contracted with three transverse projectors, but we omitted them
in order to simplify the notation. In most cases, transversality
of $V_{\alpha\beta\gamma}$ with respect to all its indices will be
restored automatically owing to the contraction with bare propagators.
The latter in the frequency-momentum
($\omega,\k$) representation have the form:
\begin{eqnarray}
\langle B_{\alpha} (\omega,\k) B'_{\beta}(-\omega,-\k) \rangle _0 &=&
\langle
B'_{\alpha}(\omega,\k) B_{\beta}(-\omega,-\k) \rangle _0^*=
\frac{1} {(-{\rm i}\omega +\kappa _0 k^2)}\,P_{\alpha\beta}(\k) ,
\nonumber \\
\langle B_{\alpha}(\omega,\k) B_{\beta}(-\omega,-\k)  \rangle _0
&=& (\BO \cdot\k)^{2}
\langle B_{\alpha}(\omega,\k) B'_{\alpha'} (-\omega,-\k) \rangle _0
\, \langle v_{\alpha'}(\omega,\k) v_{\beta'} (-\omega,-\k) \rangle _0
\langle B'_{\beta'}(\omega,\k) B_{\beta} (-\omega,-\k) \rangle _0,
\nonumber \\
\langle B_{\alpha}(\omega,\k) v_{\beta} (-\omega,-\k)\rangle _0 &=&
(\BO\cdot \k) \langle B_{\alpha}(\omega,\k)
B'_{\alpha'}(-\omega,-\k)\rangle _0
\langle v_{\alpha'}(\omega,\k) v_{\beta}(-\omega,-\k)  \rangle _0,
\nonumber \\
\langle B'_{\alpha}(\omega,\k) B'_{\beta}(-\omega,-\k) \rangle _0&=&0 ,
\label{lines}
\end{eqnarray}
and the bare propagator $\langle v_{\alpha} v_{\beta}  \rangle _0$
is given by Eqs. (\ref{2-point-v}).

The magnitude $B^{o}\equiv|\BO|$
can be eliminated from the action (\ref{action}) by rescaling
of the fields: $\B\to B^{o}\B$, $\BP\to\BP/B^{o}$.
Therefore, any total or connected Green function
of the form $\langle\B(x_{1})\cdots\B(x_{n})\, \BP(y_{1})
\cdots\BP(y_{p})\rangle$ contains the factor of $(B^{o})^{n-p}$.
The parameter $B^{o}$ appears in the bare propagators (\ref{lines})
only in the numerators. It then follows that the Green functions
with $n-p<0$ vanish identically. On the contrary, the 1-irreducible
function $\langle\B(x_{1})\cdots\B(x_{n})\, \BP(y_{1})
\cdots\BP(y_{p})\rangle_{\rm 1-ir}$ contains a factor of
$(B^{o})^{p-n}$ and therefore vanishes for $n-p>0$; this
fact will be relevant in the analysis of the renormalizability of
the model (see Sec.~\ref{sec:RG}).

The white-in-time character of $\bbox{v}$ permits to exploit the
Gaussian integration by parts
(a comprehensive description of this techniques can be found, e.g.,
in Ref.~\cite{F95}) to obtain closed, exact equations for
the equal-time correlation functions of the field $\B$.
This strategy has been used in Ref.~\cite{LM99}.
Below we
give an alternative derivation of the equation for the pair
correlation functions based on the field theoretical formulation
of the problem (see also Ref.~\cite{AA98} for the scalar case).

The pair correlation functions $\langle\Phi\Phi\rangle$ of the
multicomponent field $\Phi$ satisfy standard Dyson equation,
which in the component notation reduces to the system of two
nontrivial equations for the exact correlation function
$C_{\alpha\beta}(\omega,\k)=\langle B_{\alpha}(\omega,\k)
B_{\beta}(-\omega,-\k)  \rangle$ and the exact response function
$G_{\alpha\beta}(\omega,\k)=\langle B_{\alpha}(\omega,\k)
B'_{\beta}(-\omega,-\k)  \rangle$. The latter is independent of
$\BO$ (see above) and thus can be written as
$G_{\alpha\beta}(\omega,\k)= P_{\alpha\beta}(\k) G(\omega,k)$
with certain isotropic scalar function $G(\omega,k)$.
In our model these equations, usually referred to as the Dyson--Wyld
equations (see, e.g., Ref.~\cite{Monin}), have the form
\begin{mathletters}
\label{Dyson}
\begin{equation}
G^{-1}(\omega, k) P_{\alpha\beta}(\k)
= \bigl[-{\rm i}\omega +\kappa_0 k^{2}\bigr] P_{\alpha\beta}(\k) -
\Sigma_{\alpha\beta}^{B'B} (\omega, \k),
\label{Dyson1}
\end{equation}
\begin{equation}
C_{\alpha\beta}(\omega, \k)= |G(\omega, k)|^{2}\, \Bigl[
(\BO \cdot\k)^{2}
\langle v_{\alpha}(\omega,\k) v_{\beta} (-\omega,-\k) \rangle _0
+\Sigma _{\alpha\beta}^{B'B'} (\omega, \k)\Bigr],
\label{Dyson2}
\end{equation}
\end{mathletters}
where $\langle B_{\alpha} B_{\beta} \rangle _0$ is given in Eq.
(\ref{lines}), $\Sigma^{B'B}$ and $\Sigma^{B'B'}$ are self-energy
operators represented by the corresponding 1-irreducible diagrams;
the other functions $\Sigma^{\Phi\Phi}$ vanish identically. It is
also convenient to contract Eq. (\ref{Dyson1}) with the projector
$P_{\alpha\beta}(\k)$ in order to obtain the scalar equation:
\begin{mathletters}
\label{Dyson'}
\begin{equation}
G^{-1}(\omega, k) = -{\rm i}\omega +\kappa_0 k^{2} -
\Sigma^{B'B} (\omega, k),
\label{Dyson3}
\end{equation}
where we have written
\begin{equation}
\Sigma^{B'B} (\omega, k) \equiv
\Sigma_{\alpha\beta}^{B'B} (\omega, \k) P_{\alpha\beta}(\k) / (d-1).
\label{Dyson4}
\end{equation}
\end{mathletters}

The feature characteristic of the rapid-change models like
(\ref{action}) is that all the skeleton multiloop diagrams entering
into the self-energy operators $\Sigma^{B'B}$ and $\Sigma^{B'B'}$
contain effectively closed circuits of retarded propagators
$\langle\B\BP\rangle$ and therefore vanish; it is also crucial here
that the propagator $\langle {\bv}{\bv}\rangle$ in Eq.
(\ref{temporal}) is proportional to the $\delta$ function in time.
Therefore the self-energy operators in (\ref{Dyson}) are given by the
one-loop approximation exactly and have the form
\begin{mathletters}
\label{Dyson9}
\begin{equation}
\Sigma^{B'B} = \put(0.00,-56.00){\makebox{\dS}} \hskip1.7cm ,
\label{Dyson7}
\end{equation}
\begin{equation}
\Sigma^{B'B'} = \put(0.00,-56.00){\makebox{\dSS}} \hskip1.7cm .
\label{Dyson8}
\end{equation}
\end{mathletters}
The solid lines in the diagrams denote the exact propagators
$\langle\B\BP\rangle$ and $\langle\B\B\rangle$; the ends with a slash
correspond to the field $\BP$, and the ends without a slash correspond
to $\B$; the dashed lines denote the velocity propagator
(\ref{2-point-v}); the vertices correspond to the factor (\ref{vertex}).
The analytic expressions for the diagrams in Eq. (\ref{Dyson9}) have
the form
\begin{mathletters}
\label{sigma}
\begin{eqnarray}
\Sigma^{B'B} (\omega, k) = \frac{P_{\alpha\beta}(\k)}{(d-1)}
\int\frac{d\omega'}{2\pi} \int\frac{d{\bf q}}{(2\pi)^{d}} \,
V_{\alpha\alpha_{3}\alpha_{1}} (\k, \p, \q)\,
P_{\alpha_{3}\alpha_{4}}(\p) G(\omega',\p) \,
\frac{D_{0}\,P_{\alpha_{1}\alpha_{2}}(\q)} {q^{d+\eps}}\,
V_{\alpha_{4}\beta\alpha_{2}} (-\k, -\p, -\q) ,
\label{sigma1}
\end{eqnarray}
\begin{eqnarray}
\Sigma^{B'B'}_{\alpha\beta} (\omega, \k)=\int\frac{d\omega'}{2\pi}
\int\frac{d{\bf q}}{(2\pi)^{d}} \,
V_{\alpha\alpha_{3}\alpha_{1}} (\k, \p, \q)\,
C_{\alpha_{3}\alpha_{4}} (\omega',\p) \,
\frac{D_{0}\,P_{\alpha_{1}\alpha_{2}}(\q)} {q^{d+\eps}}\,
V_{\beta\alpha_{4}\alpha_{2}} (-\k, -\p, -\q) ,
\label{sigma2}
\end{eqnarray}
\end{mathletters}
where $ \k+\q+\p=0$, the vertex $V_{\alpha\beta\gamma}$ is defined
in Eq. (\ref{vertex}), and the explicit form (\ref{2-point-v})
of the velocity covariance is used. We also recall that
the integrations over $\q$ should be cut off from below at $q=m$.

The integrations over $\omega'$ in the right hand sides of
Eqs. (\ref{sigma}) give the equal-time
response function $G(q)=(1/{2\pi})\int{d\omega'}\,G(\omega',q)$
and the equal-time pair correlation function $C_{\alpha\beta}(\q)=
(1/{2\pi})\int{d\omega'}\,C_{\alpha\beta}(\omega',\q)$;
note that both the self-energy operators are in fact independent of
$\omega$. The only contribution to $G(q)$ comes from the bare
propagator (\ref{lines}), which in the $t$ representation is
discontinuous at coincident times. Since the correlation function
(\ref{temporal}), which enters into the one-loop diagram
for $\Sigma ^{B'B}$, is symmetric in $t$ and $t'$, the response
function must be defined at $t=t'$ by half the sum of the limits.
This is equivalent to the convention
\[G(q)=(1/{2\pi})\int{d\omega'}\,(-i\omega'+\kappa _0 q^2)^{-1}=1/2 \]
and gives
\begin{equation}
\Sigma^{B'B} (\omega, k) = \frac{P_{\alpha\beta}(\k)}{2(d-1)}
\int\frac{d{\bf q}}{(2\pi)^{d}}
V_{\alpha\alpha_{3}\alpha_{1}} (\k, \p, \q)\,
P_{\alpha_{3}\alpha_{4}}(\p)  \,
\frac{D_{0}\,P_{\alpha_{1}\alpha_{2}}(\q)} {q^{d+\eps}}\,
V_{\alpha_{4}\beta\alpha_{2}} (-\k, -\p, -\q) .
\label{sigma22}
\end{equation}
Substituting Eq. (\ref{vertex}) into Eq. (\ref{sigma22}) after
lengthy but straightforward calculation gives
\begin{equation}
\Sigma^{B'B} (\omega, k) = (-1/2) k_{\alpha} k_{\beta} D_{0}\,
\int\frac{d{\bf q}}{(2\pi)^{d}}
\frac{P_{\alpha_{1}\alpha_{2}}(\q)} {q^{d+\eps}}.
\label{sigma3}
\end{equation}
The integration over ${\bf q}$ in Eq. (\ref{sigma3}) is performed
explicitly using the relation
\begin{eqnarray}
\int d{\bf q}\, f(q)\frac{q_{i}q_{j}}{q^{2}}  =
\frac{\delta_{ij}}{d} \int d{\bf q}\, f(q)
\label{isotropy}
\end{eqnarray}
and gives
\begin{mathletters}
\label{otvet}
\begin{equation}
\Sigma^{B'B} (\omega, k)= -k^{2}\,
\frac{D_{0}\,(d-1)}{2d} \,  J(m),
\label{otvet1}
\end{equation}
where we have written
\begin{equation}
J(m)\equiv\int\frac{d{\bf q}}{(2\pi)^{d}}\,
\frac{1}{q^{d+\eps}} = C_{d}\, m^{-\eps} /\eps.
\label{otvet2}
\end{equation}
\end{mathletters}
Here and below $C_{d} \equiv  {S_{d}}/{(2\pi)^{d}}$ and
$S_d\equiv 2\pi ^{d/2}/\Gamma (d/2)$ is the surface area of the
unit sphere in $d$-dimensional space; the parameter $m$ has
arisen from the lower limit in the integral over $\q$.

Equations (\ref{Dyson'}), (\ref{otvet}) give an explicit exact expression
for the response function in our model; it will be used in Sec.
\ref{sec:RG}
for the exact calculation of the RG functions. Like in the scalar case,
the exact response function differs from its bare analog (\ref{lines})
simply by the substitution $\kappa_{0} \to \kappa_{0} + D_{0}\,(d-1)\,
J(m) / 2d$. Below we use the intermediate
expression (\ref{sigma3}). The integration of Eq. (\ref{Dyson2})  over
the frequency $\omega$ gives a closed equation for the equal-time
correlation function; it is important here that the $\omega$ dependence of
the right hand side is contained only in the prefactor $|G(\omega,
k)|^{2}$.
Using Eq. (\ref{sigma3}) the equation for $C_{\alpha\beta} (\k)$
can be written in the form
\begin{eqnarray}
2\bigl(\kappa_{0} k^{2}+ \Sigma^{B'B}\bigr) C_{\alpha\beta} (\k) =
(\BO \cdot\k)^{2}
\langle v_{\alpha}(\omega,\k) v_{\beta} (-\omega,-\k) \rangle _0
\nonumber \\
+ \int\frac{d{\bf q}}{(2\pi)^{d}}
V_{\alpha\alpha_{3}\alpha_{1}} (\k, \p, \q)\,
C_{\alpha_{3}\alpha_{4}} (\p)
\frac{D_{0}\,P_{\alpha_{1}\alpha_{2}}(\q)} {q^{d+\eps}}\,
V_{\beta\alpha_{4}\alpha_{2}} (-\k, -\p, -\q) ,
\label{9}
\end{eqnarray}
and using Eqs. (\ref{vertex}) and (\ref{sigma3}) it can be rewritten as
\begin{eqnarray}
& {} &
2\kappa_{0} k^{2} C_{\alpha\beta} (\k) = (\BO \cdot\k)^{2}
\langle v_{\alpha}(\omega,\k) v_{\beta} (-\omega,-\k) \rangle _0
\nonumber \\ &+&
\int\frac{d{\bf q}}{(2\pi)^{d}}
\frac{D_{0}} {q^{d+\eps}}\, \Bigl\{
q_{\alpha_{1}}q_{\alpha_{2}}
C_{\alpha_{1}\alpha_{2}}(\p) P_{\alpha\beta} (\q) -
p_{\alpha_{1}} q _{\alpha_{2}} C_{\alpha\alpha_{2}}(\p)
P_{\alpha_{1}\beta} (\q) -
p_{\alpha_{2}} q _{\alpha_{1}}
C_{\alpha_{1}\beta}(\p) P_{\alpha\alpha_{2}} (\q) \Bigr\}
\nonumber \\ &+&
\int\frac{d{\bf q}}{(2\pi)^{d}}
\frac{D_{0}\, P_{\alpha_{1}\alpha_{2}} (\q)} {q^{d+\eps}}\, \Bigl\{
p_{\alpha_{1}}p_{\alpha_{2}} C_{\alpha\beta}(\p)  -
k_{\alpha_{1}}k_{\alpha_{2}} C_{\alpha\beta}(\k) \Bigr\}.
\label{10}
\end{eqnarray}
For $0<\eps <2$, Eq. (\ref{10}) allows for the limit $m\to0$: the
first three integrals in its right hand side are separately finite
for $m=0$; the last integral is finite owing to the subtraction,
which has come from the contribution with $\Sigma^{B'B}$ in the
left hand side of Eq. (\ref{9}). Indeed, the possible IR
divergence of this integral at ${\bf q}=0$ is suppressed by the
vanishing of the expression in the curly brackets.
In what follows we set $m=0$.

Equation (\ref{10}) can also be rewritten as a partial differential
equation for the pair correlation function in the coordinate
representation,
$C_{\alpha\beta}(\bbox{r})\equiv\langle B_{\alpha}
(t,\bbox{x}) B_{\beta}(t,\bbox{x+r})\rangle$
[we use the same notation $C_{\alpha\beta}$ for the coordinate
function and its Fourier transform].
Noting that the integrals in Eq. (\ref{10}) involve convolutions of
the functions $C_{\alpha\beta}(\k)$ and $D_{0}P_{\alpha\beta}(\k)/
k^{d+\eps}$, the Fourier transform of the spatial part
(\ref{spatial}) of the velocity correlation function
(\ref{2-point-v}),
and replacing the momenta by the corresponding derivatives,
${\rm i}p_{\alpha} \to \partial_{\alpha}$ and so on, we obtain:
\begin{eqnarray}
2\kappa_{0} \partial^2 C_{\alpha\beta} = -
\left (\partial_{\alpha_{1}}\partial_{\alpha_{2}} S_{\alpha\beta}\right )
\left (B_{\alpha_{1}}^o B_{\alpha_{2}}^o+  C_{{\alpha_{1}}{\alpha_{2}}}
\right) +
\left (\partial_{\alpha_{2}}\,S_{\alpha_{1}\beta}\right )
\partial_{\alpha_{1}}\,C_{\alpha \alpha_{2}}
 \nonumber \\ +
\left (\partial_{\alpha_{1}}S_{\alpha {\alpha_{2}}}\right )\left
(\partial_{\alpha_{2}} C_{\alpha_{1}\beta}\right )-
S_{\alpha_{1}\alpha_{2}}\,\partial_{\alpha_{1}}\partial_{\alpha_{2}} \,
C_{\alpha\beta}.
\label{eq-moto}
\end{eqnarray}
Note that the correlation function (\ref{2-point-v}) enters into
Eq. (\ref{eq-moto})
only through the function $S_{\alpha_{1}\alpha_{2}}$ from Eq.
(\ref{eddydiff}), or, in other words, through the difference
(\ref{difference}), which has a finite limit at $m=0$. The $m$ dependent
constant part of (\ref{spatial}) vanishes under the differentiation
in the first four terms in the right hand side of Eq. (\ref{eq-moto}),
and in the last term it is subtracted explicitly, owing
to the subtraction in Eq. (\ref{10}). Equation (\ref{eq-moto})
should be augmented by the solenoidality condition:
\begin{equation}
\partial_{\alpha} C_{\alpha\beta} = 0 .
\label{cont}
\end{equation}
For the nonstationary state, the function
$C_{\alpha\beta}(t,\bbox{r})\equiv\langle B_{\alpha}
(t,\bbox{x}) B_{\beta}(t,\bbox{x+r})\rangle$
depends explicitly on $t$, and the term $\partial_t\, C_{\alpha\beta}$
appears on the right hand side of Eq. (\ref{eq-moto}),
see, e.g., Ref. \cite{LM99}.

 \section{Nonperturbative results for the scaling exponents of the
 2-point magnetic correlation function} \label{sec:2-point}

In this section we focus our attention on the inertial-range behavior
of the second-order equal-time correlation function
$C_{\alpha\beta}(t,\bbox{r})\equiv \langle B_{\alpha}(t,\bbox{r})
B_{\beta}(t,\bbox{0}) \rangle$ in the statistically steady state.
As shown in Ref.~\cite{LM99}, a steady state
is present when $\xi<1$, $\xi=1$ being the threshold of instability.
As such threshold coincides with that of the isotropic problem \cite{V96},
it follows that dynamo effect is thus not swithched on by anisotropic
contributions.

In the isotropic case, the analytic expression for the scaling exponent
of $C_{\alpha\beta}$ has been obtained in Ref.~\cite{V96}.
It was also shown by the author of \cite{V96} that the anomalous
exponent is universal, and the anomaly is associated with
zero-mode solutions of the equations satisfyed by $C_{\alpha\beta}$.
Higher-order correlation function exponents have been calculated to
$O(\xi)$ in Ref.~\cite{AA98} by exploiting the RG.

With respect to Ref.~\cite{V96}, the main technical difference is that,
in order  to ex\-tract the ani\-so\-tro\-pic contributions
to the isotropic scaling, the angular structure of zero modes has now
to be explicitly taken into account.

To start our analysis, let us consider the closed Eq.~(\ref{eq-moto})
for $C_{\alpha\beta}$. For what follows, it is worth emphasizing
two properties of $C_{\alpha\beta}$:

\noindent (i) because of
homogeneity, $C_{\alpha\beta}$ is left invariant under the
following set of transformations:
\begin{eqnarray}
\bbox{r}\longmapsto -\bbox{r}\qquad&\mbox{and}&\qquad
\alpha  \longleftrightarrow \beta \label{simme1};
\label{simme2}
\end{eqnarray}
(ii) $C_{\alpha\beta}(\bbox{r})=C_{\alpha\beta}(-\bbox{r})$, as it follows
from (\ref{eq-moto}) after the substitution $\bbox{r}\mapsto -\bbox{r}$.

In the presence of anisotropy, the most general
expression for the two-point magnetic correlations,
$C_{\alpha\beta}(\bbox{r})$, in the stationary state involves five
(two in the isotropic case) functions depending on both $r\equiv
|\bbox{x}-\bbox{x}'|$ and $z\equiv
\cos\theta=\hat{\bbox{B}}^o\cdot\bbox{r}/r$, where $\hat{\bbox{B}}^o$
is the unit vector corresponding to the direction selected by the mean
magnetic field.
Remark that the space is anisotropic but still homogeneous,
so that there is no explicit dependence on the points $
\bbox{x},\bbox{x}'$,
but only on their difference.
 Namely,
\begin{equation}
C_{\alpha\beta}(\bbox{r})
= {\cal F}_1(r,z)\; \frac{r_{\alpha}r_{\beta}}{r^2}+
{\cal F}_2(r,z)\delta_{\alpha\beta}+ {\cal F}_3(r,z)
\frac{\hat{B}^o_{\alpha} r_{\beta}}{r}+
{\cal F}_4(r,z)\frac{\hat{B}^o_{\beta} r_{\alpha}}{r}+
 {\cal F}_5(r,z)  \hat{B}^o_{\alpha}\;\hat{B}^o_{\beta}.
\label{general}
\end{equation}
 From the properties i) and ii) of $C_{\alpha\beta}(\bbox{r})$ one
immediately obtains
the following relations for the ${\cal F}$'s:
\begin{eqnarray}
{\cal F}_i(r,z)&=&{\cal F}_i(r,-z)\qquad i=1,2,5 \label{rela}\\
{\cal F}_3(r,z)&=&-{\cal F}_3(r,-z)\label{reld}\\
{\cal F}_3(r,z)&=&{\cal F}_4(r,z)\label{rele}.
\end{eqnarray}
Substituting the expression (\ref{general}) into (\ref{eq-moto}) and using
the
chain rules, we obtain, after lengthy but straightforward algebra, the
following four equations
(corresponding to the projections over
$r_{\alpha}r_{\beta}/r^2$, $\delta_{\alpha\beta}$,
$\hat{B}^o_{\alpha} r_{\beta}/r$
and $\hat{B}^o_{\alpha}\;\hat{B}^o_{\beta}$):
\begin{eqnarray}
%
%
&[&a_1r^2\partial^2_r+b_1 r\partial_r +
c_1 (1-z^2)\partial^2_z+
d_1 z\partial_z +e_1] {\cal F}_1+\nonumber\\
&[&f_1 r\partial_r+g_1 z\partial_z + j_1]\,{\cal F}_2 +
[k_1 z\, r\partial_r +l_1z^2\partial_z + m_1z +n_1\partial_z]{\cal
F}_3+\nonumber\\
&[&o_1 +p_1z^2]\,{\cal F}_5 = (q_1+r_1z^2)\,B^{o\;2}\label{eq1}\\
&&\vspace{-10mm}\nonumber\\
%
%
&a&_2 {\cal F}_1+[b_2r^2\partial^2_r+c_2 r\partial_r +
d_2 (1-z^2)\partial^2_z+
e_2 z\partial_z +f_2]\, {\cal F}_2+\nonumber\\
&g&_2 z\,{\cal F}_3+
\left[k_2 +l_2z^2 \right]\,{\cal F}_5 = (m_2+n_2z^2)\,B^{o\;2}\label{eq2}
\\
&&\vspace{-10mm}\nonumber\\
&a&_3 \partial_z  {\cal F}_1+b_3\partial_z  {\cal F}_2+\nonumber\\
&[&c_3r^2\partial^2_r + d_3 r\partial_r + e_3 (1-z^2)\partial^2_z+
f_3 z\partial_z + g_3 ]\,{\cal F}_3+\nonumber\\
&[&j_3 z\,r\partial_r+(k_3+l_3z^2)\partial_z+m_3 z]\,{\cal F}_5
= n_3\,B^{o\;2}\, z\label{eq3}\\
&&\vspace{-10mm}\nonumber\\
%
%
&a&_4\partial_z  {\cal F}_3+ \nonumber\\
&[&b_4 r^2\partial^2_r + c_4 r\partial_r + d_4 (1-z^2)\partial^2_z+
e_4 z\partial_z+f_4]\,{\cal F}_5=g_4\, B^{o\;2},\label{eq4}
\end{eqnarray}
where the coefficients $a_i, b_i, \cdots r_i$ are functions of
$\xi$ and $d$
and are reported in Appendix \ref{blocco1_coeff}. Without loss
of generality, we have fixed $D=1$ in (\ref{eddydiff}),
and we have neglected all terms involving
the magnetic diffusivity $\kappa_0$, our attention being indeed focused in
the
inertial range of scales, i.e.,~$\eta \ll r\ll L$
where $\eta=\kappa_0^{1/\xi}\propto \Lambda^{-1}$
is the dissipative scale for the problem.

With the substitution of the expression (\ref{general}),
the solenoidal condition (\ref{cont}) splits into
the following couple of equations:
\begin{eqnarray}
&[&r\partial_r +(d-1)]\, {\cal F}_1 +[r\partial_r
-z\partial_z]\, {\cal F}_2+\nonumber\\
&[&z\,r\partial_r +\partial_z -
z^2\partial_z -z]\, {\cal F}_3=0\label{eqcont1}\\
&&\vspace{-10mm}\nonumber\\
&\partial&_z {\cal F}_2+ [r\partial_r + d]\,{\cal F}_3+
[z\,r\partial_r + (1-z^2)\partial_z]\,{\cal F}_5 = 0\,, \label{eqcont2}
\end{eqnarray}
associated to the projections over $r_{\beta}/r$ and $\hat{B}^o_{\beta}$,
respectively.

 From the relation (\ref{rele}), Eqs.~(\ref{eqcont1}) and (\ref{eqcont2})
it then follows
that only two functions of the ${\cal F}'s$ in (\ref{general}) are
independent.
 A possible way to isolate contributions
of the anisotropic components
from the isotropic scaling is to use the decomposition of
${\cal F}'s$ on the Legendre polynomial basis.
This is the subject of the next subsection.

\subsection{Decomposition in Legendre polynomials}
\label{decompo}
In terms of the Legendre polynomials, functions ${\cal F}_i(r,z)$
can be decomposed in the form:
\begin{eqnarray}
&{\cal F}&_i(r,z)=\sum_{j=0}^{\infty} f_{j}^{(i)}(r)\,
P_{j}(z)\qquad i=1,2,5\qquad \mbox{(j even)}
\label{f125}\\
&{\cal F}&_3(r,z)=\sum_{j=0}^{\infty} f_{j}^{(3)}(r)\, P_{j}(z)
\qquad\mbox{(j odd)},
\label{f3}
\end{eqnarray}
where the separation of even and odd orders in (\ref{f125}) and (\ref{f3})
arises as a consequence of the symmetries expressed by the relations
(\ref{rela}) and (\ref{reld}), respectively.

Simple considerations related to the `uniaxial' character of the
forcing term with $\bbox{B}^o$ in the basic equation (\ref{fp}),
and the linearity of the latter in $\bbox{B}^o$ and $\bbox{B}$
suggest that the index $j$ in the above decompositions should
be restricted to $j\le 2$. The rigorous assessment of this point
will be given  in Subsec.~\ref{admiss}. On the other hand, contributions
associated to $j>2$ can be easily ``activated'' either when a fully
anisotropic forcing (i.e.,~projecting onto all Legendre polynomials)
is added on the right hand side  of (\ref{fp}), or in the framework
of finite-size systems led by anisotropic boundary conditions.
Moreover, as we shall see in Sec.~\ref{sec:RG}, scaling exponents
associated to $j>2$ contribute to the inertial-range scaling of
higher-order correlation functions involving the product
$B_{\alpha}B_{\beta}$ at a single spacetime point.
The latter property holds also without the invocation of a fully
anisotropic forcing on the left hand side of Eq.~(\ref{fp}).
 From all these considerations we shall exploit the general
decompositions (\ref{f125}) and (\ref{f3}) involving all $j$'s.

In order to obtain equations for $f_{j}^{(i)}(r)$, we have
to insert Eqs.~(\ref{f125}) and (\ref{f3}) into
Eqs.~(\ref{eq1})-(\ref{eqcont2}).
Furthermore, quantities like  $z^p\partial_z^q{\cal F}$ ($p=0,1,2$ and
$q=0,1$)
and $(1-z^2)\partial_z^2 {\cal F}$ have to be expressed in terms
of Legendre polynomials. This can be done exploiting well-known
relations involving the Legendre polynomials (see, e.g.,
Ref.~\cite{Grad65}):
the resulting expressions for $z^p\partial_z^q{\cal F}$ and
$(1-z^2)\partial_z^2 {\cal F}$ are reported in Appendix
\ref{appa_legendre}.
For the sake
of brevity, we report hereafter only the projection of equation
(\ref{eq1}),
the structure of the others  being indeed similar (the full set of
equations
is however reported in Appendix \ref{appa-eq-lunghe}):

\begin{eqnarray}
%
%
&a_1&r^2 f_j^{''(1)}
+b_1 r f_j^{'(1)}  + c_1 \left[j(1-j) f_j^{(1)} +
2(2j+1)\sum_{q=1}^{\infty} f_{2q+j}^{(1)}
\right] + d_1 \left[ j f_j^{(1)} + (2j+1)
\sum_{q=1}^{\infty} f_{2q+j}^{(1)}
\right] +e_1 f_j^{(1)} +\nonumber\\
&f_1& r f_j^{'(2)} +g_1 \left[ j f_j^{(2)} + (2j+1)
\sum_{q=1}^{\infty} f_{2q+j}^{(2)}
\right]+ j_1  f_{j}^{(2)} +
k_1 r \left[\frac{j}{2j-1}f_{j-1}^{'(3)}
+\frac{j+1}{2j+3}f_{j+1}^{'(3)}\right] +
n_1 (2j+1)\sum_{q=0}^{\infty}f_{2q+j+1}^{(3)}+\nonumber\\
&l_1& \left[ \frac{j(j-1)}{2j-1}f_{j-1}^{(3)}-\frac{(j+1)(j+2)}{2j+3}
f_{j+1}^{(3)} + (2j+1)\sum_{q=0}^{\infty}f_{2q+j+1}^{(3)} \right]
+ m_1 \left[ \frac{j}{2j-1}f_{j-1}^{(3)} + \frac{j+1}{2j+3}f_{j+1}^{(3)}
\right] +\nonumber\\
&o_1& f_{j}^{(5)} + p_1 \left[ \frac{j(j-1)}{(2j-1)(2j-3)}f_{j-2}^{(5)} +
\left(\frac{(j+1)^2}{(2j+3)(2j+1)} + \frac{j^2}{(2j-1)(2j+1)}\right)
f_{j}^{(5)} + \frac{(j+2)(j+1)}{(2j+3)(2j+5)}f_{j+2}^{(5)}
\right]= \nonumber\\
&B&^{o\;2}
\left[q_1+r_1 \left(\frac{2}{3}\delta_{j,2}+\frac{1}{3}\delta_{j,0}\right)
\right].\label{eq1pro}
\end{eqnarray}

 From the above equation we can see that terms like
$z^p\partial_z^q{\cal F}$ are responsible for the coupling
between an arbitrary
anisotropic contribution of order $j$ and all larger orders.
The full set of equations (\ref{eq1proap})--(\ref{eqcont2proap})
is thus not closed and there are no chances to solve them analytically.
Simple physical argumentations actually permit to overcome the closure
problem. Indeed, in the presence of a cascade-like mechanism
of energy transfer towards small scales,
anisotropy present at the
integral scale should rapidly decay during the multiple-step transfer, and
an almost isotropic inertial range scaling should be restored.
Mathematically, this means that $f_j^{(i)}$'s should
be rapidly decreasing functions of the degree of anisotropy $j$, i.e.,
\begin{equation}
f_1^{(i)} << f_2^{(i)} << \cdots
\label{super-sim}
\end{equation}
and similarly for their derivatives.
We shall  control the validity of this physical
assumption in a self-consistent way, at the end of
our calculation.\footnote{The physical assumption (\ref{super-sim})
is unnecessary when the decomposition in the irreducible representations
of the $SO(d)$ symmetry group is exploited (see Ref.~\cite{ABP99} for
the case $d=3$). This leads  exactly to the same results obtained earlier
in Ref.~\cite{LM99}, where the hierarchy (\ref{super-sim}) was assumed.
Notice that the additional exponents (subsets II and III) reported in
\cite{ABP99} are related to the pseudotensorial structures and that in our
model
they do not contribute to the inertial-range behavior of the pair
correlator.}

The hierarchy (\ref{super-sim}) is exploited here by retaining,
for each $i$ appearing in the functions $f_p^{(i)}$'s, the lowest
value of the index $p$. When doing this, the simplifications
on Eqs.~(\ref{eq1proap})-(\ref{eqcont2proap}) are enormous and the
resulting
set of equations reads:

\begin{eqnarray}
%
%
&a_1&r^2 f_j^{''(1)}
+b_1 r f_j^{'(1)}  + \left[ c_1 j(1-j)
+ d_1  j  +e_1 \right] f_j^{(1)} +
f_1 r f_j^{'(2)} + \left[g_1 j + j_1 \right] f_{j}^{(2)} +
k_1 r \frac{j}{2j-1}f_{j-1}^{'(3)} + \nonumber\\
&&\hspace{-3mm}\left[ l_1 \frac{j(j-1)}{2j-1}
+ m_1 \frac{j}{2j-1}\right] f_{j-1}^{(3)}  +
p_1  \frac{j(j-1)}{(2j-1)(2j-3)}f_{j-2}^{(5)} = B^{o\;2}
\left[q_1+r_1 \left(\frac{2}{3}\delta_{j,2}+\frac{1}{3}\delta_{j,0}\right)
\right]\label{neweq1}\\
\vspace{-1mm}\nonumber\\
%
%
&a_2& f_{j}^{(1)}+ b_2 r^2 f_{j}^{''(2)}
+c_2 r f_{j}^{'(2)}+ \left[
d_2 j(1-j) +
e_2 j  +f_2\right] f_j^{(2)}+g_2
\frac{j}{2j-1}f_{j-1}^{(3)}+\nonumber\\
&l_2& \frac{j(j-1)}{(2j-1)(2j-3)}f_{j-2}^{(5)} = B^{o\;2}
\left[m_2+n_1 \left(\frac{2}{3}\delta_{j,2}+\frac{1}{3}\delta_{j,0}\right)
\right]\label{neweq2} \\
&&\vspace{-1mm}\nonumber\\
&a_3& (2j+1) f_{j+1}^{(1)}+b_3 (2j+1) f_{j+1}^{(2)}+
c_3 r^2 f_{j}^{''(3)}+ d_3 r f_{j}^{'(3)}+
\left[e_3 j(1-j)+ f_3 j
+ g_3 \right]f_j^{(3)}+\nonumber\\
&j_3& r
\frac{j}{2j-1}f_{j-1}^{'(5)}  +
\left[l_3 \frac{j(j-1)}{2j-1}
+m_3 \frac{j}{2j-1}\right] f_{j-1}^{(5)}
= n_3\,B^{o\;2}\, \delta_{j,2}\label{neweq3}\\
&&\vspace{-1mm}\nonumber\\
%
%
&a_4& (2j+1) f_{j+1}^{(3)}
+ b_4 r^2 f_{j}^{''(5)}+ c_4 r f_{j}^{'(5)}+
\left[ d_4  j(1-j) + e_4 j +f_4 \right]f_{j}^{(5)}=
g_4\, B^{o\;2}\delta_{j,0}\label{neweq4}\\
&&\vspace{-1mm}\nonumber\\
%
%
&r&f_{j}^{'(1)} +(d-1)f_{j}^{(1)} + r f_{j}^{'(2)}
-j f_j^{(2)}
+ r \frac{j}{2j-1}f_{j-1}^{'(3)}- \frac{j^2}{2j-1} f_{j-1}^{(3)}
=0\label{neweqcont1}\\
%
%
&&\hspace{-3mm}(2j+1) f_{j+1}^{(2)}+
r f_{j}^{'(3)} + d\, f_{j}^{(3)}+ r \frac{j}{2j-1}f_{j-1}^{'(5)}
-\frac{j(j-1)}{2j-1}f_{j-1}^{(5)}= 0 \label{neweqcont2}.
\end{eqnarray}

Some remarks are worth. Focusing on the isotropic contribution, $j=0$,
we notice that the first two equations involve solely the functions
$f_{0}^{(1)}$ and $f_{0}^{(2)}$ (and their derivatives). With the
solenoidal
condition (\ref{neweqcont1}), it is easy to check that they coincide with
the equation reported in Ref.~\cite{V96}
for the isotropic problem. Moreover, for
$j\geq 2$, Eq.~(\ref{neweq1}) suggests to take the function
$\bbox{f}_j\equiv
(f_{j}^{(1)}, f_{j}^{(2)}, f_{j-1}^{(3)}, f_{j-2}^{(5)})$
as an unknown field. It is immediately verified that $\bbox{f}_j$
appears also in the other equations when the index $j$ ($j\geq2$)
is renamed $(j-1)$ in Eqs.~(\ref{neweq3}) and (\ref{neweqcont2})
and $(j-2)$ in Eq.~(\ref{neweq4}). When doing this, a linear
partial differential equation
(PDE) system of the type ${\cal L}_j \bbox{f}_j = \bbox{g}_j$
(hereafter, repeated indices are not summed)
is obtained, $\bbox{g}_j$ involving all terms related to the
mean field $B^{o}$, and ${\cal L}_j$ is restricted, for instance,
to the first four equations.

The analytical treatment
of the resulting equation system still remains a very hard
task for general values of the space separation $r$. The situation
changes when one focuses on the inertial range of scales
(i.e.,~for $\eta \ll r\ll L$). In the latter case scaling  behaviors
are expected and we shall have:
\begin{equation}
f_j^{(i)}(r) \propto
r^{\zeta_j^{(i)}}\qquad\mbox{with}\qquad\zeta_0^{(i)}<\zeta_1^{(i)}<\cdots
\label{decay}
\end{equation}
where the hierarchy on the exponents  $\zeta_j^{(i)}$ immediately follows
from  (\ref{super-sim}).

The structure of the above equations fixes the relation between the scaling
exponents relative to different $f$'s. Indeed, when searching for
power law  solutions $f^{(i)}_j(r)\propto r^{\zeta_j^{(i)}}$,
in order to obtain  balanced equations the `oblique'
relations must hold:
\begin{equation}
\zeta_j\equiv
\zeta_j^{(1)}=\zeta_j^{(2)}=\zeta_{j-1}^{(3)}=\zeta_{j-2}^{(5)}.
\label{obliqua}
\end{equation}

We are now ready to show that nontrivial scaling behaviors for
$\bbox{f}_j$
take place due to zero modes, i.e.,~solutions of the homogeneous problem
${\cal L}_j \bbox{f}_j =0$.
To that purpose, we exploit (\ref{obliqua}) and define coefficients
$\bbox{y}_j$ through the relation $\bbox{f}_j \equiv \bbox{y}_j
r^{\zeta_j}$.
Inserting the latter expression in the PDE system, a $4\times 4 $
algebraic linear system for $\bbox{y}_j$ is obtained.
The emergence of
zero modes is thus reduced
to impose the existence of nontrivial solutions
of a $4\times 4$ homogeneous linear system, that means here the resolution
of an  algebraic equation of 8-th degree arising from the condition that
the
determinant of the system  coefficients is zero.
The calculation, lengthy but straightforward, leads
to four sets of zero modes (actually eight sets, but it turns out that
the associated coefficients, $\bbox{y}_j$, of
four of them do not satisfy the solenoidal condition
(\ref{neweqcont1})-(\ref{neweqcont2}))
the expressions and the admissibility
of which are given and discussed in the next section.

\subsection{Zero-mode solutions and their admissibility}
\label{admiss}
Let us start with the case $j=0$, corresponding to the
isotropic contribution. As already observed,
Eqs.~(\ref{neweq1}) and (\ref{neweq2})
are decoupled from the others, and the problem can be solved
directly for $f_0^{(1)}$ and $f_0^{(2)}$ which must satisfy also
the solenoidal condition (\ref{neweqcont1}). The imposition of
the existence of nontrivial solutions (for $j=0$ we have a homogeneous
$2\times 2 $
algebraic linear system for $\bbox{y}_j$) and the solenoidal
constraint (\ref{neweqcont1}) lead to the following solutions:

\begin{equation}
\zeta_0^{\pm}=\frac{
-d^2+d-2\xi\pm\sqrt{12d^2\xi-8d\xi+8\xi^2d -4d^2\xi^2-4d^3\xi
+d^2+d^4-2d^3}}{2d-2}
\end{equation}
with their $\xi\to 0$ and $d\to \infty$ limits:
\begin{eqnarray}
\zeta_0^{+}&=& -\xi+O(\xi^2) \\
           &=& -\xi -\frac{2\xi^2}{d}+O(1/d^2) \nonumber\\
\zeta_0^{-}&=& -d+\xi \frac{d-3}{d-1} + O(\xi^2)\\
           &=&-\frac{1}{d} + \frac{2\xi(\xi-1)}{d}+ O(1/d^2).\nonumber
\end{eqnarray}

For $j\geq2$ zero-mode exponents are:

\begin{eqnarray}
\zeta_j^{\pm} &=&  - \frac {1}{2(d-1)} \Big\{2\,\xi + d^{2} - d -
\Big [ - 2\,d^{3}\,\xi  - 2\,d^{2}\,\xi^{2} -\nonumber\\
&6&\hspace{-1mm}d^{3} + 4\, \xi^{2}\,d + 8
+ 10\,d\,
\xi  + 20\,d\,j - 20\,d - 8\,\xi  - \nonumber\\
&8&\hspace{-1mm} j + 4\,d^{2}\,j^{
2} + 2\,\xi^{2}- 4\,\xi\,j^{2} + 17\,d^{2} - 8\,d\,j^{2} + \nonumber\\
&8&\hspace{-1mm} \xi \,j + 4\,d^{3}\,j + 4\,d^{2}\,j\,\xi
 + 4\,d\,j^{2}\,\xi  + 4\,j^{2} - 16\,d^{2}\,j -\nonumber\\
&1&\hspace{-2mm}2\,d\,\xi\,j + d^{4} \pm
\left . \left . 2\sqrt{K} \left( d-1\right) \left( 2-\xi\right)
\right]^{1/2}\right\}
\label{amore}
\end{eqnarray}

\begin{eqnarray}
\rho_j^{\pm} &=&  - \frac {1}{2(d-1)} \Big\{2\,\xi + d^{2} - d +
\Big[ - 2\,d^{3}\,\xi  - 2\,d^{2}\,\xi^{2} -\nonumber\\
&6&\hspace{-1mm}d^{3} + 4\, \xi^{2}\,d + 8
+ 10\,d\,
\xi  + 20\,d\,j - 20\,d - 8\,\xi  - \nonumber\\
&8&\hspace{-1mm} j + 4\,d^{2}\,j^{
2} + 2\,\xi^{2}- 4\,\xi\,j^{2} + 17\,d^{2} - 8\,d\,j^{2} + \nonumber\\
&8&\hspace{-1mm} \xi \,j + 4\,d^{3}\,j + 4\,d^{2}\,j\,\xi
 + 4\,d\,j^{2}\,\xi  + 4\,j^{2} - 16\,d^{2}\,j -\nonumber\\
&1&\hspace{-2mm}2\,d\,\xi\,j + d^{4} \pm
\left . \left . 2\sqrt{K} \left( d-1\right) \left( 2-\xi\right)
\right]^{1/2}\right\}
\label{odio}
\end{eqnarray}
where
\begin{eqnarray}
 K &=& (d - 1)\,(d^{3} + 4\,d^{2}\,j - 5\,d^{2} + 2\,d^{
2}\,\xi + \xi^{2}\,d + 4\,d\,\xi
\,j - \nonumber\\
&6&\hspace{-0.6mm}d\,\xi  + 8\,d - 12\,d\,j + 4\,d\,j^{2} - \xi^{2} +
4\,\xi
 + 8\,j - 8\,\xi \,j - \nonumber\\
&4&\hspace{-1.4mm} - 4\,j^{2} + 4\,\xi \,j^{2})\nonumber
\end{eqnarray}
with their $\xi\to 0$ and $d\to \infty$ limits:
\begin{eqnarray}
\zeta_j^+&=&
j-\xi\frac{(d-1+j)(d^2+dj-2d+4j-2j^2)}{(d-2+2j)(d+2j)(d-1)}+
O(\xi^2)\\
&=& j-\xi +\frac{2\xi(j-\xi)}{d} +O(1/d^2)\nonumber\\
\zeta_j^-&=& j-2+\xi \frac{-4d^2-17dj+16d +28j-16
-14j^2+2d^2j+5dj^2+2j^3}{(d-2+2j)(d+2j-4)(d-1)}+O(\xi^2)\\
&=& j-2+\frac{2\xi(j-2)}{d}+O(1/d^2)\nonumber\\
\rho_j^+&=&
-d-j+\xi\frac{-5d^2-7d j +6d +4 j-2j^2+d^3+2d^2j -dj^2-2j^3}
{(d-2+2j)(d+2j)(d-1)}+ O(\xi^2)\\
&=&-\frac{1}{d}+\xi-j-\frac{2\xi(j+1-\xi)}{d} +O(1/d^2)\nonumber\\
\rho_j^-&=& 2-d-j-\xi\frac{(j-1)(2j^2-4j+5dj-4d+2d^2)}
{(d-2+2j)(d+2j-4)(d-1)} +O(\xi^2)\\
&=& -\frac{1}{d}+ 2-j -\frac{2\xi(j-1)}{d} +O(1/d^2).\nonumber
\end{eqnarray}

Let us discuss the admissibility of these solutions. With the term
admissible we mean a solution $\bbox{f}_j(r)$
satisfying the appropriate boundary conditions,
at both small (UV limit) and large scales (IR limit).
Specifically, the following asymptotic behaviors have to be satisfied:
\begin{equation}
\bbox{f}_j(r)\qquad\mbox{regular for}\qquad r\sim\eta\to 0
\label{UV}
\end{equation}
\begin{equation}
\bbox{f}_j(r) \to 0\qquad\mbox{for}\qquad r>>L.
\label{IR}
\end{equation}

Concerning the  limit (\ref{UV}), we have to consider solutions
corresponding to the diffusive range and to match them with
our inertial range power laws. From Eq.~(\ref{eq-moto}) we can easily see
that the equations holding in the diffusive range are obtained by setting
to zero the parameter $\xi$. The consequence is that our
inertial range zero-mode solutions become solutions in the diffusive
range for $\xi=0$.
The problem  related to the UV boundary condition
is thus reduced to search for regular solutions for
$\bbox{f}_j(r)$ in the $\xi \to 0$ limit. This is easy to do and the result
is that solely exponents $\zeta_0^+$
and $\zeta_j^{\pm}$ for $j\geq 2$ permit to
satisfy the condition of regularity for $\bbox{f}_j$,
the other exponents  being indeed $\leq 0$ for $\xi=0$.
Notice that,
the zero-mode exponent $\zeta_0^+$ coincides with the isotropic
solution obtained in Ref.~\cite{V96}.

Let us pass to discuss the IR boundary conditions (\ref{IR}).
In this case,
as pointed out in Ref.~\cite{V96},
a crucial role is played by the external forcing. Indeed, in the presence
of forcing, zero modes and the decaying forced solution may be matched
at the integral scale $L$, thus satisfying the IR boundary conditions.
The result of this argument (which can be rigorously illustrated
solely for $j=0$ where  the general solution
for $\bbox{f}_j(r)$ is available) is that zero-mode exponents are not
admissible for $j\geq 4$. Indeed, as we can see from
Eq.~(\ref{neweq1})--(\ref{neweq4}), the forcing term related to $B^{o}$
projects solely on the shells $j\leq 2$.

To summarize, we have one admissible zero mode for $j=0$ ($\zeta_0^+$)
and two admissible zero modes for $j=2$ ($\zeta_2^{\pm}$). Our
attention being focused on the inertial range of scales (i.e.,~$r/L<<1$),
our choice for $j=2$ is for $\zeta_2^{-}$. We have indeed to take
the exponent giving the leading inertial range contribution.

Finally, we can thus define the final solution $\zeta_j$ of our problem as:

\begin{eqnarray}
\zeta_0\equiv \zeta_0^{+}&=&\frac{
-d^2+d-2\xi + \sqrt{12d^2\xi-8d\xi+8\xi^2d -4d^2\xi^2-4d^3\xi
+d^2+d^4-2d^3}}{2d-2}\nonumber\\
&=& -\xi+O(\xi^2) \nonumber\\
           &=& -\xi -\frac{2\xi^2}{d}+O(1/d^2) \label{expo1}\\
\zeta_2\equiv \zeta_2^{-} &=& -\frac{1}{2(d-1)}
\Big\{d^2-d+2\xi-\Big[8-12d-8\xi+2d\xi+4\xi^2d-2d^2\xi^2-2d^3\xi+8d^2\xi+
d^2+\nonumber\\
&&\left.\left.
2\xi^2+2d^3+d^4-2\sqrt{K}(2-\xi)(d-1)\right]^{1/2}\right\}   \nonumber\\
&=&\frac{2\xi}{(d-1)(d+2)}+ O(\xi^2)\nonumber\\
&=& \frac{2\xi}{d^2}+O(1/d^3)\label{expo2}
\end{eqnarray}
where
\[
K=  (d-1)(d^3+2d^2\xi+3d^2+\xi^2d+2d\xi-4+4\xi-\xi^2).
\]
We stress that, for $j>2$,
exponents $\zeta_j\equiv \zeta_j^{-}$  become admissible
under the conditions already discussed in Subsec.~\ref{decompo}.

\begin{figure}
 \centerline{\psfig{file=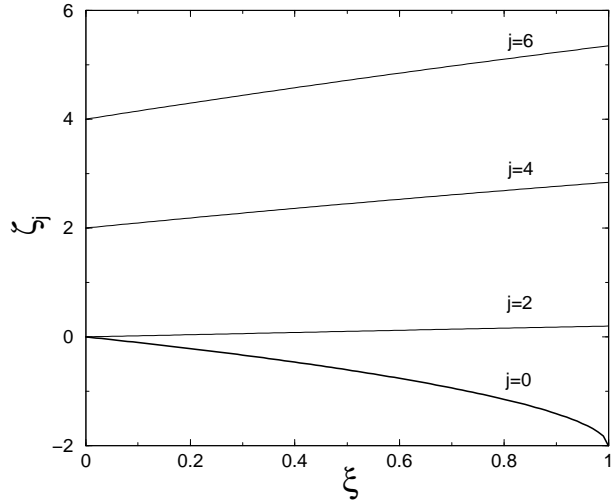,width=8cm}}
 \vspace{1mm}
\caption{Behavior of zero-mode exponents
$\zeta_j$ ($j=0,2,4$ and $6$)
{\it vs} $\xi$ for $d=3$.
Notice the inequality  $\zeta_0<\zeta_2<\zeta_4\cdots$ that
means the validity of the hierarchy (\protect\ref{decay}) and thus the
self-consistency of our zero-mode solutions.}
\label{fig1}
\end{figure}

The last remark concerns the self-consistency of our solution,
that is, the validity of the hierarchy in (\ref{decay}).
The validity of the latter can
be easily verified from Fig.~\ref{fig1} where the
behavior of $\zeta_j$ ($j=0,2,4$ and $6$)
are shown for $d=3$ as a function of $\xi$.
Similar behaviors actually hold for all $d$'s and $j$'s.
As we shall see in Sec.~\ref{sec:Operators}, a hierarchical order
for the scaling exponents is also present for higher-order correlation
functions.

 \section{UV renormalization of the model. RG functions and RG equations}
 \label {sec:RG}

The RG approach to the statistical models of the turbulence is
exposed in Refs. \cite{UFN,turbo} in detail
(see also Ref.~\cite{RG} for the scalar Kraichnan model);
below we confine ourselves to the only information we need.

The analysis of UV divergences is based on the analysis of
canonical dimensions (see, e.g., Ref.~\cite{Zinn}). Dynamical models
of the type (\ref{action}), in contrast to static models, have
two scales, i.e., the canonical (``engineering'')  dimension of
some quantity $F$ (a field or a parameter in the action functional)
is described by two numbers, the momentum dimension $d_{F}^{k}$ and
the frequency dimension $d_{F}^{\omega}$. They are determined so that
$[F] \sim [\ell]^{-d_{F}^{k}} [T]^{-d_{F}^{\omega}}$, where $\ell$
is the length scale and $T$ is the time scale. The dimensions are
found from the obvious normalization conditions
\[ d_k^k=-d_{\bf x}^k=1,\quad d_k^{\omega}
=d_{\bf x}^{\omega }=0, \quad d_{\omega }^k=d_t^k=0, \quad
d_{\omega }^{\omega }=-d_t^{\omega }=1, \]
and from the requirement
that each term of the action functional be dimensionless (with
respect to the momentum and frequency dimensions separately).
Then, based on $d_{F}^{k}$ and $d_{F}^{\omega}$,
one can introduce the total canonical dimension
$d_{F}=d_{F}^{k}+2d_{F}^{\omega}$ (in the free theory,
$\partial_{t}\propto\partial^{2}$), which plays in the theory of
renormalization of dynamical models the same role as the
conventional (momentum) dimension does in static problems.

In the action (\ref{action}), there are fewer terms than fields
and parameters, and the canonical dimensions are not determined
unambiguously. This is of course a manifestation of the fact that
the ``superfluous'' parameter $B^{o}\equiv|\BO|$ can be scaled out
from the action (see Sec.~\ref{sec:FT}). After it has been eliminated
(or, equivalently, zero canonical dimensions have been assigned
to it), the definite canonical dimensions can be assigned to the
other quantities. They are given in Table \ref{table1},
including the dimensions of renormalized parameters, which will
appear later on. From Table \ref{table1} it follows that the model
becomes logarithmic (the coupling constant $g_{0}$ becomes
dimensionless) at $\eps=0$, and the UV divergences
have the form of the poles in $\eps$ in the Green functions.

The total canonical dimension of an arbitrary
1-irreducible Green function $\Gamma = \langle\Phi \cdots \Phi
\rangle _{\rm 1-ir}$ is given by the relation
\begin{equation}
d_{\Gamma }=d_{\Gamma }^k+2d_{\Gamma }^{\omega }=
d+2-N_{\Phi }d_{\Phi},
\label{deltac}
\end{equation}
where $N_{\Phi}=\{N_{\BP},N_{\B},N_{\bv}\}$ are the
numbers of corresponding fields $\Phi\equiv\{\BP, \B, {\bv}\}$
entering into the function $\Gamma$, and the summation over all
types of the fields is implied. The total dimension $d_{\Gamma}$
is the formal index of the UV divergence. This means that
superficial UV divergences, whose removal requires counterterms,
can be present only in those functions $\Gamma$ for which
$d_{\Gamma}$ is a non-negative integer.

Analysis of divergences in the problem (\ref{action}) should be based on
the following auxiliary considerations:

(i) All the 1-irreducible Green functions with
$N_{\BP}< N_{\B}$ vanish (see Sec.~\ref{sec:FT}).

(ii) If for some reason a  number of external momenta occur as an
overall factor in all the diagrams of a given Green function, the
real index of divergence $d_{\Gamma}'$ is smaller than $d_{\Gamma}$
by the corresponding number (the Green function requires
counterterms only if $d_{\Gamma}'$  is a non-negative integer).

In the model (\ref{action}), the derivative $\partt$ at the vertex
can be moved onto the field $\BP$ using the integration
by parts, which decreases the real index of divergence:
$d_{\Gamma}' = d_{\Gamma}- N_{\BP}$.
The field $\BP$ enters into the counterterms only
in the form of a derivative, $\partial_{\alpha}B_{\beta}'$.

(iii) A great deal of diagrams in the model (\ref{action}) contain
effectively closed circuits of retarded propagators
$\langle\B\BP\rangle_{0}$  and therefore vanish. For example, all
the nontrivial diagrams of the 1-irreducible function
$\langle B_{\alpha} B'_{\beta}{v}_{\gamma}\rangle_{\rm 1-ir}$ vanish.

 From the dimensions in Table \ref{table1} we find
$d_{\Gamma} = d+2 - N_{\bv} - dN_{\BP}$
and $d_{\Gamma}'=(d+2) - N_{\bv} - (d+1) N_{\BP}$.
 From these expressions it follows that for any $d$,
superficial divergences can only exist in the 1-irreducible functions
with
$N_{\BP}=1$,  $N_{\bv}=N_{\B}=0 $
($d_{\Gamma}=2$, $d_{\Gamma}'=1$),
$N_{\BP}=N_{\B}=1$, $N_{\bv}=0$
($d_{\Gamma}=2$, $d_{\Gamma}'=1$),
$N_{\BP}=N_{\bv}=1$, $N_{\B}=0$
($d_{\Gamma}=1$, $d_{\Gamma}'=0$), and
$N_{\BP}=N_{\B}=N_{\bv}=1$,
($d_{\Gamma}=1$, $d_{\Gamma}'=0$)
[we recall that $ N_{\B}\le N_{\BP}$ (see (i) above)].
However, the first of these counterterms has necessarily the form
of a total derivative, $B^{o}_{\alpha}\partial^{2}B_{\alpha}$,
vanishes after the integration over ${\bfx}$ and therefore gives no
contribution to the renormalized action. Furthermore, for the
last two of these functions, all the nontrivial diagrams vanish
[see (iii) above]. As in the case of the passive scalar field
\cite{RG}, we are left with the only superficially divergent
function $\langle B_{\alpha}'B_{\beta}\rangle_{\rm 1-ir}$;
the corresponding counterterm necessarily contains the derivative
$\partial $ and therefore reduces to $B_{\alpha}'\partial^{2}B_{\alpha}$
(another structure,
$B_{\alpha}'\partial_{\alpha}\partial_{\beta} B_{\beta}$,
vanishes by virtue of the solenoidality of $\B$).

Introduction of this counterterm is reproduced by the multiplicative
renormalization of the parameters $g_{0},\kappa_0$ in the action
functional (\ref{action}) with the only
independent renormalization constant $Z_{\kappa}$:
\begin{equation}
\kappa_0=\kappa Z_{\kappa},\quad g_{0}=g\mu^{\eps}Z_{g}, \quad
Z_{g}=Z_{\kappa}^{-1}.
\label{reno}
\end{equation}
Here $\mu$ is the reference mass in the minimal subtraction
scheme (MS), which we always use in what follows, $g$ and $\kappa$
are renormalized analogs of the bare parameters $g_{0}$ and $\kappa_0$,
and $Z=Z(g,\eps,d)$ are the renormalization constants. Their
relation in Eq. (\ref{reno}) results from the absence of renormalization
of the contribution with $K^{-1}$ in Eq. (\ref{action}), so that
$D_{0}\equiv g_{0}\kappa_0 = g\mu^{\eps} \kappa$. No renormalization
of the fields and the ``mass'' $m$ is required,
i.e., $Z_{\Phi}=1$ for all $\Phi$ and $m_{0}=m$.
The renormalized action functional has the form
\begin{equation}
S_{R}(\Phi)= \BP \Bigl[ - \partial_{t}\B -({\bv}\cdot\partt) \B +
(\B\cdot\partt){\bv}  + (\BO\cdot\partt){\bv} +
 Z_{\kappa}\kappa \partial^{2} \B
\Bigr] -{\bv} K^{-1} {\bv}/2,
\label{renact}
\end{equation}
where the function $K$ from Eq. (\ref{spatial}) is
expressed in renormalized parameters using Eqs. (\ref{reno}):
$D_{0} =g_{0}\kappa_{0} =g  \mu^\eps \kappa$.

The relation $ S(\Phi,e_{0})=S_{R}(\Phi,e,\mu)$ (where
$e_{0}=\{g_{0}, \kappa_{0}, m\} $ is the complete set of bare
parameters, and $e =\{g, \kappa, m\}$ is the set of renormalized
parameters) implies $W(e_{0})=W_{R}(e,\mu)$ for the bare
correlation functions $W=\langle \Phi\cdots \Phi \rangle$
and their renormalized analogs $W_{R}$.
We use $\widetilde{\cal D}_{\mu}$
to denote the differential operation $\mu\partial_{\mu}$ for fixed
$e_{0}$ and operate on both sides of this equation with it. This
gives the basic RG differential equation:
\begin{equation}
{\cal D}_{RG}\,W_{R}(e,\mu) = 0,
\label{RG1}
\end{equation}
where ${\cal D}_{RG}$ is the operation $\widetilde{\cal D}_{\mu}$
expressed in the renormalized variables:
\begin{equation}
{\cal D}_{RG}\equiv {\cal D}_{\mu} + \beta(g)\partial_{g}
-\gamma_{\kappa}(g){\cal D}_{\kappa}.
\label{RG2}
\end{equation}
In Eq. (\ref{RG2}), we have written ${\cal D}_{x}\equiv x\partial_{x}$ for
any variable $x$, and the RG functions (the $\beta$ function and
the anomalous dimension $\gamma$) are defined as
\begin{mathletters}
\label{RGF}
\begin{equation}
\gamma_{F}(g)\equiv \Dm \ln Z_{F}\qquad  {\rm for\ any\ } Z_{F},
\label{RGF1}
\end{equation}
\begin{equation}
\beta(g)\equiv\Dm  g=g[-\eps+\gamma_{\kappa}(g)].
\label{RGF2}
\end{equation}
\end{mathletters}
The relation between $\beta$ and $\gamma$ in Eq. (\ref{RGF2})
results from the definitions and the last relation in Eq. (\ref{reno}).

Now let us turn to the explicit calculation of the constant
$Z_{\kappa}$ in the one-loop approximation in the MS scheme.
It is determined by the requirement that the 1-irreducible function
$\langle\BP\B\rangle_{{\rm 1-ir}}$ expressed in renormalized
variables be UV finite (i.e., be finite for $\eps\to0$). This
requirement determines $Z_{\kappa}$ up to an UV finite contribution; the
latter is fixed by the choice of a renormalization scheme. In the MS
scheme all renormalization constants have the form ``1 + only poles
in  $\eps$.''  The function $\langle\BP\B\rangle_{{\rm 1-ir}}$
in our model is
known exactly [see Eqs.~(\ref{Dyson1}) and (\ref{otvet})].  Let us
substitute Eqs. (\ref{reno}) into Eqs. (\ref{Dyson1}), (\ref{otvet})
and choose $Z_{\kappa}$ to cancel the pole in $\eps$ in the integral
$J(m)$. This gives:
\begin{equation}
Z_{\kappa} = 1 - g\, C_{d} \frac{(d-1)}{2d\eps},
\label{Z}
\end{equation}
with the coefficient $C_{d}$ from Eq. (\ref{otvet2}).
Note that the result (\ref{Z}) is exact,
i.e., it has no corrections of order $g^{2}$, $g^{3}$, and so on;
this is a consequence of the fact that the one-loop approximation
(\ref{otvet}) for the response function is exact. Note also that
Eq. (\ref{Z}) coincides literally with the exact expression for
$Z_{\kappa}$ in the case of a passive scalar (see Ref.~\cite{RG}).

For the anomalous dimension $\gamma_{\kappa}(g)\equiv
\widetilde{\cal D}_\mu \ln Z_{\kappa}
=\beta(g)\partial_{g}\ln Z_{\kappa}$ from the relations (\ref{RGF2})
and (\ref{Z}) one obtains:
\begin{equation}
\gamma_{\kappa}(g)=\frac{-\eps {\cal D}_g
\ln Z_{\kappa}}{1-{\cal D}_g \ln Z_{\kappa}}=
g\, C_{d} \frac{(d-1)}{2d}.
\label{gammanu}
\end{equation}
 From Eq. (\ref{RGF2}) it then follows that the RG equations of the model
have an IR stable fixed point [$\beta(g_{*})=0$, $\beta'(g_{*})>0$]
with the coordinate
\begin{equation}
g_{*}= \frac{2d\eps}{C_{d}(d-1)}.
\label{fixed}
\end{equation}

Let $F(r)$ be some equal-time two-point quantity, for example,
the pair correlation function of the primary fields
$\Phi\equiv\{\BP, \B,{\bv}\}$  or some composite operators.
We assume that $F(r)$ is multiplicatively renormalizable,
i.e., $F=Z_{F}F^{R}$ with certain renormalization constant $Z_{F}$.
The existence of nontrivial IR stable
fixed point implies that in the IR asymptotic region $\Lambda r>>1$
and any fixed $mr$ the function $F(r)$ takes on the form
\begin{equation}
F(r) \simeq  \kappa_{0}^{d_{F}^{\omega}}\, \Lambda^{d_{F}}
(\Lambda r)^{-\Delta_{F}}\, \chi(mr),
\label{RGR}
\end{equation}
where $d_{F}^{\omega}$ and $d_{F}$ are the frequency and total
canonical dimensions of $F$, respectively, and $\chi$ is some
function whose explicit form is not determined by
the RG equation itself. The critical dimension $\Delta_{F}$
is given by the expression
\begin{equation}
\Delta[F]\equiv\Delta_{F} = d_{F}^{k}+ \Delta_{\omega}
d_{F}^{\omega}+\gamma_{F}^{*},
\label{32B}
\end{equation}
where $\gamma_{F}^{*}$ is the value of the anomalous dimension
(\ref{RGF1}) at the fixed point and
$\Delta_{\omega}=2-\gamma^{*}_{\kappa}=2-\eps$ is the critical dimension
of frequency [note that the value
of $\gamma_{\kappa}(g)$ at the fixed point is also found  exactly
from the last relation in Eq. (\ref{RGF2}):
$\gamma_{\kappa}^{*} \equiv \gamma_{\kappa}(g_*)= \eps$].

The critical dimensions of the basic fields $\Phi$
in our model are found exactly [we recall that they are
not renormalized and therefore $\gamma_{\Phi}=0$ for all $\Phi$].
 From the dimensions in Table \ref{table1} we then find
$\Delta_{\bv}=1-\eps$, $\Delta_{\theta} = 0$,
$\Delta_{\theta'} = d$.

 \section{Critical dimensions of composite operators}
 \label {sec:Operators}

Any local (unless stated to be otherwise) monomial or polynomial
constructed of primary fields and their derivatives at a single
spacetime point $x\equiv \{t,{\bfx}\}$ is termed a composite operator.
Examples are $\B^2$, $B_{\alpha}' \partial^{2} B_{\beta}$,
$B_\alpha'\bbox{v}\cdot\bbox{\partial}B_\alpha$, and so on.

Since the arguments of the fields coincide, correlation functions with
these operators contain additional UV divergences, which are removed
by additional renormalization procedure (see,
e.g., Ref.~\cite{Zinn}). For the renorma\-li\-zed correlation functions
standard RG equations are obtained, which describe IR scaling with
definite critical dimensions $\Delta_{F}\equiv\Delta[F]$  of
certain ``basis'' operators  $F$. Owing to the renormalization,
$\Delta[F]$ does not coincide in general with the naive sum of
critical dimensions of the fields and derivatives entering into $F$.

In general, composite operators are mixed in renormalization,
i.e., an UV finite renormalized operator $F^{R}$ has the form
$F^{R}=F+$ counterterms, where the contribution of the
counterterms is a linear combination of $F$ itself and,
possibly, other unrenormalized operators which ``admix''
to $F$ in renormalization.

Let $F\equiv\{F_{a}\}$ be a closed set, all of whose monomials mix
only with each other in renormalization. The renormalization matrix
$Z_{F}\equiv\{Z_{ab}\}$ and the  matrix of anomalous dimensions
$\gamma_{F}\equiv\{\gamma_{ab}\}$ for this set are given by
\begin{equation}
F_{a}=\sum _{b} Z_{ab}
F_{b}^{R},\qquad \gamma _F=Z_{F}^{-1}\D_{\mu }Z_{F},
\label{2.2}
\end{equation}
and the corresponding matrix of critical dimensions
$\Delta_{F}\equiv\{\Delta_{ab}\}$ is given by Eq. (\ref{32B}),
in which $d_{F}^{k}$, $d_{F}^{\omega}$, and $d_{F}$ are understood as
the diagonal matrices of canonical dimensions of the operators in
question (with the diagonal elements equal to sums of corresponding
dimensions of all fields and derivatives constituting $F$) and
$\gamma^{*}_{F}\equiv\gamma_{F} (g_{*})$ is the
matrix (\ref{2.2}) at the fixed point (\ref{fixed}).

Critical dimensions of the set $F\equiv\{F_{a}\}$ are
given by the eigenvalues of the matrix $\Delta_{F}$. The ``basis''
operators that possess definite critical dimensions have the form
\begin{equation}
F^{bas}_{a}=\sum_{b} U_{ab}F^{R}_{b}\ ,
\label{2.5}
\end{equation}
where the matrix $ U_{F} =  \{U_{ab} \}$ is such that
$\Delta'_{F}= U_{F} \Delta_{F} U_{F}^{-1}$ is diagonal.

In general, counterterms to a given operator $F$ are
determined by all possible 1-irreducible Green functions
with one operator $F$ and arbitrary number of primary fields,
$\Gamma=\langle F(x) \Phi(x_{1})\cdots\Phi(x_{n})\rangle_{\rm 1-ir}$.
The total canonical dimension (formal index of divergence)
for such functions is given by
\begin{equation}
d_\Gamma = d_{F} - N_{\Phi}d_{\Phi},
\label{index}
\end{equation}
with the summation over all types of fields
$\Phi\equiv\{\BP, \B,{\bv}\}$ entering into the function.
For superficially divergent diagrams, the real index
of divergence, $d_\Gamma'=d_\Gamma-N_{\BP}$, is a non-negative
integer, cf. Sec. \ref{sec:RG}.

In what follows, an important role will be played by the
tensor composite operators built solely of the field $\B$
without derivatives:
\begin{equation}
F^{(np)}_{\alpha_{1}\cdots \alpha_{p}} (x) \equiv
B_{\alpha_{1}}(x) \cdots B_{\alpha_{p}}(x) \,
[B_{\alpha}(x)B_{\alpha}(x)]^{l},  \qquad n\equiv 2l+p.
\label{Fnp}
\end{equation}
Here $p$ is the rank of the tensor and $n= 2l+p$ is the total number
of fields $\B$ entering into the operator.
 From Table I and Eq. (\ref{index}) for the operators (\ref{Fnp})
we obtain $d_{F}=0$ and $d_\Gamma' = -N_{\bv} -(d+1)N_{\BP}$.
Therefore, the divergences can exist only in the functions
with $N_{\bv} =N_{\BP}=0$, for which $d_\Gamma=d_\Gamma'=0$.
This means that the operators $F^{(np)}$ mix only with each other,
i.e., the set (\ref{Fnp}) is closed with respect to the
renormalization.

The simple analysis of the diagrams shows that the
1-irreducible function
\begin{equation}
\langle F^{(np)}(x) \B(x_{1})\cdots\B(x_{n'})\rangle_{\rm 1-ir}
\label{razryv}
\end{equation}
contains the factor $(B^{o})^{n-n'}$ and therefore vanishes
for $n'>n$, cf. the discussion in the end of Sec. \ref{sec:FT}.
It then follows that the operator $F^{(n'p')}$ can admix to
$F^{(np)}$ only if $n'\le n$. This means that
the corresponding infinite renormalization matrix
\begin{equation}
F^{(np)} = \sum_{n'p'} \, Z_{np,n'p'} \,F_{R}^{(n'p')}
\label{Matrix}
\end{equation}
is in fact block-triangular, i.e., $Z_{np,n'p'} =0$ for $n'> n$,
and so are the matrices $\gamma_{F}$, $\Delta_{F}$ and $U_{F}$.
It is then obvious that the critical dimensions associated with the
operators $F^{(np)}$  are completely determined by the eigenvalues of the
finite subblocks with $n'= n$. In the following, we shall not be interested
in the precise form of the basis operators (\ref{2.5}), we rather shall
be interested in the anomalous dimensions themselves. Therefore, we
can neglect all the elements of the matrix (\ref{Matrix}) other than
$Z_{np,np'}$. The latter are found from the functions
(\ref{razryv}) which are independent of $\BO$ and therefore can be
calculated directly in the isotropic theory with $\BO=0$. It is then
clear that the block $Z_{np,np'}$ can be
diagonalized by the changing to irreducible operators: scalars
($p=0$), vectors ($p=1$) and traceless tensors ($p\ge2$),
but for our purposes it is sufficient to note that the elements
of the block $Z_{np,np'}$ vanish for $p<p'$, i.e., this
block is triangular along with the corresponding blocks
of the matrices $\gamma_{F}$, $\Delta_{F}$ and $U_{F}$.
Indeed, the irreducible tensor of the rank $p$ consists of the
monomials with $p'\le p$ only, for example,
$F^{(22)}_{\alpha\beta} = B_{\alpha}B_{\beta} - \delta_{\alpha\beta}
\B^2 /d$, and therefore only these monomials can admix to
the monomial of the rank $p$ in renormalization.
The final conclusion is that the critical dimensions, associated
with the set (\ref{Fnp}), coincide with the diagonal elements
$\Delta_{n,p}\equiv\Delta_{np,np}$ of the matrix (\ref{32B}),
they are completely determined by the diagonal elements
$Z_{np}\equiv Z_{np,np}$ of the matrix (\ref{Matrix}),
and that they can be calculated directly in the isotropic
theory with $\BO=0$.

Now let us turn to the one-loop calculation of the diagonal element
$Z_{np}$ of the matrix $Z_{F}$ in the MS scheme.
Let $\Gamma(x;\B)$ be the generating functional of the
1-irreducible Green functions with one composite operator $F^{(np)}$
and any number of fields $\B$. Here $x\equiv
\{t,{\bfx}\}$ is the argument of the operator and $\B(x)$ is
the functional argument, the ``classical analog'' of the random
field $\B(x)$. We are interested in the $n$-th term of the
expansion of $\Gamma(x;\B)$ in $\B(x)$, which we denote
$\Gamma^{(n)}(x;\B)$; it has the form
\begin{eqnarray}
\Gamma^{(n)}_{\alpha_{1}\cdots \alpha_{p}}(x;\B) = \frac{1}{n!} \,
\sum_{\beta_{1}\cdots\beta_{n}}\, \int dx_{1} \cdots \int dx_{n}\,
B_{\beta_{1}}(x_{1})\cdots B_{\beta_{n}}(x_{n}) \
\langle F^{(np)}_{\alpha_{1}\cdots \alpha_{p}}(x)
B_{\beta_{1}}(x_{1})\cdots
B_{\beta_{n}}(x_{n})\rangle_{\rm 1-ir}.
\label{Gamma1}
\end{eqnarray}
In the one-loop approximation the functional (\ref{Gamma1}) is
represented diagrammatically as follows:
\begin{equation}
\Gamma^{(n)}_{\alpha_{1}\cdots \alpha_{p}}=
F^{(np)}_{\alpha_{1}\cdots \alpha_{p}} +\frac{1}{2}
\put(-20.00,-50.00){\makebox{\dA}} \hskip1.4cm .
\label{Gamma2}
\end{equation}
Here the solid lines denote the {\it bare} propagators
$\langle \B\B\rangle _{0}$ from Eq. (\ref{lines}), the ends with
a slash correspond to the field $\BP$, and the ends without a slash
correspond to $\B$; the dashed line denotes the velocity propagator
(\ref{2-point-v}); the vertices correspond to the factor (\ref{vertex}).
The first term in Eq. (\ref{Gamma2}) is the ``tree'' approximation, and
the black circle with two attached lines in the diagram denotes the
variational derivative
\begin{equation}
 V_{\alpha_{1}\cdots \alpha_{p}\,\beta_{1}\beta_{2}} (x;\, x_{1}, x_{2})
\equiv \frac {\ \delta^{2} F^{(np)}_{\alpha_{1}\cdots \alpha_{p}}(x)}
{\delta B_{\beta_{1}}(x_{1})\delta B_{\beta_{2}}(x_{2})}.
\label{suspect}
\end{equation}
It is convenient to represent it in the form
\begin{equation}
V_{\alpha_{1}\cdots \alpha_{p}\,\beta_{1}\beta_{2}}(x;\, x_{1}, x_{2})=
\delta(x-x_{1})\, \delta(x-x_{2})\, \frac{\partial^{2}}{\partial
b_{\beta_{1}} \partial b_{\beta_{2}}}\, \Bigl[ b_{\alpha_{1}}\cdots
b_{\alpha_{p}}\, (b^{2})^{l} \Bigr],
\label{Vertex}
\end{equation}
where $b_{\alpha}$ is a constant vector, which {\it after the
differentiation} is substituted with the field $B_{\alpha}(x)$.

The vertex (\ref{Vertex}) contains $(n-2)$ factors of $\B$.
Two remaining ``tails'' $\B$  are attached to the lower vertices
of the diagram in Eq.
(\ref{Gamma2}). We know that the UV divergent part of the diagram
is proportional to $n$ factors $\B$ {\it without derivatives},
so that we can omit the first term of the vertex
$ \BP \left[-({\bv}\cdot\partt) \B + (\B\cdot\partt){\bv} \right]$,
or, equivalently, the first term in Eq. (\ref{vertex}). Furthermore,
we can set all the external momenta in the integrand
equal to zero, and the UV divergent part of the diagram
(\ref{Gamma2}) takes on the form
\begin{equation}
b_{\beta_{3}} b_{\beta_{4}}
\frac{\partial^{2}}{\partial
b_{\beta_{1}} \partial b_{\beta_{2}}}\, \Bigl[ b_{\alpha_{1}}\cdots
b_{\alpha_{p}}\, (b^{2})^{l} \Bigr]\,
T_{\beta_{1}\beta_{2}\beta_{3}\beta_{4}},
\label{diagr01}
\end{equation}
where we have denoted
\begin{equation}
T_{\beta_{1}\beta_{2}\beta_{3}\beta_{4}} =  D_{0} \,
\int \frac{d\omega}{2\pi} \int \frac{d{\bf q}}{(2\pi)^{d}} \,
\frac{q_{\beta_{3}}q_{\beta_{4}} \,
P_{\beta_{1}\beta_{2}}({\bf q})}
{q^{d+\eps}\,[  \omega^{2}+\kappa_{0}^2 q^{4}]} \, .
\label{diagr02}
\end{equation}
We recall that the integration over $\q$ should be cut off from below
at $q=m$ (see Sec.~\ref{sec:Def}). In Eq.~(\ref{diagr02}), we have to
change to the renormalized variables using Eqs. (\ref{reno}); in
our approximation this reduces to the substitution
$g_{0}\to g\mu^{\eps}$ and $\kappa_{0}\to\kappa$.
Then we perform the integration over $\omega$ and use the relations
(\ref{isotropy}) and
\[ \int d{\bf q}\, f(q)\frac{q_{\beta_{1}}q_{\beta_{2}}q_{\beta_{3}}
q_{\beta_{4}}}{q^{4}}  =
\frac {\delta_{\beta_{1}\beta_{2}}\delta_{\beta_{3}\beta_{4}}
+\delta_{\beta_{1}\beta_{4}}\delta_{\beta_{2}\beta_{3}}+
\delta_{\beta_{1}\beta_{3}}\delta_{\beta_{2}\beta_{4}}}{d(d+2)}
\int d{\bf q}\, f(q) . \]
This gives
\begin{equation}
T_{\beta_{1}\beta_{2}\beta_{3}\beta_{4}} =
\frac{g\mu^{\eps}\, J(m)} {2d(d+2)}\,
\Bigl[(d+1)\delta_{\beta_{1}\beta_{2}} \delta_{\beta_{3}\beta_{4}}
- (\delta_{\beta_{1}\beta_{4}} \delta_{\beta_{2}\beta_{3}}
+\delta_{\beta_{1}\beta_{3}} \delta_{\beta_{2}\beta_{4}}) \Bigr],
\label{diagr2}
\end{equation}
with the integral $J(m)$ defined in Eq. (\ref{otvet2}).

Substituting Eq. (\ref{diagr2}) into Eq. (\ref{diagr01}) gives
the desired expression for the divergent part of the diagram
(\ref{Gamma2}). It is sufficient
to take into account only the terms proportional to the monomial
$B_{\alpha_{1}}(x) \cdots B_{\alpha_{p}}(x) \,
[B_{\alpha}(x)B_{\alpha}(x)]^{l}$
and neglect all the other terms, namely, those
containing the factors of $\delta_{\alpha_{1}\alpha_{2}}$ etc.
The latter determine non-diagonal elements of the matrix $Z_{F}$,
which we are not interested in here. Finally we obtain
\begin{equation}
\Gamma^{(n)}_{\alpha_{1}\cdots \alpha_{p}}\simeq
F^{(np)}_{\alpha_{1}\cdots \alpha_{p}}
\left[1 - \frac {g\mu^{\eps} \,
J(m)\,Q_{np}} {4d(d+2)}  \right] + \cdots,
\label{diagr3}
\end{equation}
where we have written
\begin{eqnarray}
Q _{np}& \equiv & 2n\,(n-1) - (d+1)\, (n-p)\, (d+n+p-2) =
\nonumber \\  &=& 2p\,(p-1) - (d-1)\, (n-p)\, (d+n+p) .
\label{Qnp}
\end{eqnarray}
The dots in Eq. (\ref{diagr3}) stand for the $O(g^{2})$ terms and
the structures different from $F^{(np)}$, $\simeq$ denotes the
equality up to UV finite parts; we also recall that $n=p+2l$.

The constant $Z_{np}$ is found from the requirement
that the renormalized analog $\Gamma_{n}^{R}\equiv
Z^{-1}_{np}\Gamma_{n}$ of the function (\ref{diagr3})
be UV finite (mind the minus sign in the exponent); along with
the expression (\ref{otvet2}) for the integral $J(m)$ and the MS
scheme this gives
\begin{equation}
Z_{np}=1- \frac{g C_{d} \,Q_{np}}{4d(d+2)\eps} +O(g^{2}),
\label{Znp}
\end{equation}
with $C_{d}$ from Eq. (\ref{otvet2}). For the anomalous dimension
$\gamma_{np}= \Dm\ln Z_{np}$ it then follows
\begin{equation}
\gamma_{np}(g)=  \frac{g C_{d} \,Q_{np}}{4d(d+2)} +O(g^{2}).
\label{Gnp}
\end{equation}
 From Table \ref{table1} and Eqs. (\ref{fixed}) and (\ref{32B})
for the corresponding critical dimension $\Delta_{n,p}=
\gamma_{np}(g_{*})$ we finally obtain
\begin{equation}
\Delta_{n,p}=\frac{\eps\,Q_{np}}{2(d-1)(d+2)}+O(\eps^{2}),
\label{Dnp}
\end{equation}
with the polynomial $Q_{np}$ from Eq. (\ref{Qnp}).

The straightforward analysis of the expression (\ref{Dnp}) shows
that for fixed $n$ and any $d$, the dimension $\Delta_{n,p}$
decreases monotonically with $p$ and reaches its minimum for
the minimal possible value, i.e., $p=0$ if $n$ is
even and $p=1$ if $n$ is odd:
\begin{mathletters}
\label{hier}
\begin{equation}
\Delta_{n,p} > \Delta_{n,p'}\qquad  {\rm if} \quad  p>p'.
\label{hier2}
\end{equation}
Furthermore, this minimal value is negative and it decreases
monotonically as $n$ increases:
\begin{equation}
0>\Delta_{2k,0}>\Delta_{2k+1,1}>\Delta_{2k+2,0}.
\label{hier3}
\end{equation}
Finally, we note that for any fixed $p$, the dimension (\ref{Dnp})
decreases monotonically as $n$ increases:
\begin{equation}
\Delta_{n,p} > \Delta_{n',p}\qquad  {\rm if} \quad  n<n'.
\label{hier4}
\end{equation}
\end{mathletters}
The inequalities (\ref{hier}) show that the critical dimensions
of the tensor operators (\ref{Fnp}) exhibit a kind of hierarchy;
in particular, the less is the rank, the more negative is the
dimension and, as will be explained in Sec. \ref{sec:OPE},
the more important is its contribution to the inertial-range behavior.

In the model of passive scalar advection by the rapid-change velocity
field (\ref{2-point-v}) in the presence of an imposed linear gradient,
similar inequalities are satisfied by the critical dimensions of
tensor operators of the type (\ref{Fnp}), but constructed of
{\it gradients} of the scalar field (see Ref.~\cite{A98}).
In the order $O(\eps)$ their critical dimensions coincide
exactly with (\ref{Dnp}), which is, however, an artifact of the
one-loop approximation (see Sec.~\ref{sec:OPE}).

As already said above, the operators that possess definite
critical dimensions (\ref{Dnp}) are not (\ref{Fnp}) themselves,
but the basis operators related to the latter by the relations
(\ref{2.2}) and (\ref{2.5}). In the isotropic case ($\BO=0$),
the basis operator with the dimension $\Delta_{n,p}$ is a
$p$-th rank traceless tensor constructed of all the monomials
$F^{(n'p')}$ with $n'= n$ and $p'\le p$. When the background
field $\BO$ is ``turned on,'' the admixture of the monomials with
$n'< n$ and $p'> p$ becomes possible. The ``missing'' fields $\B$ in the
monomials with $n'< n$ are substituted with the constant fields $\BO$
(the total number of the fields $\B$ and $\BO$ has to be equal $n$,
owing to the linearity of the basic equation (\ref{fp}) in $\B$
and $\BO$), while the ``superfluous'' indices of the monomials
with $p'> p$ are contracted with the indices of $\BO$, so that the
basis operator remains a $p$-th rank traceless tensor.
And vice versa, the unrenormalized monomial $F^{(np)}$ from (\ref{Fnp})
is a linear combination of the basis operators (\ref{2.5}) with
respective dimensions $\Delta_{n',p'}$. The hierarchy relations
(\ref{hier}) then show that the {\it minimal} dimension
entering into $F^{(np)}$ is $\Delta_{n,p_{n}}$, where
$p_{n}$ is the minimal possible value of $p$ for a given $n$,
i.e., $p_{n}=0$ if $n$ is even and $p_{n}=1$ if $n$ is odd.

 \section{Operator product expansion and the anomalous scaling for
 the correlation functions} \label {sec:OPE}

The representation (\ref{RGR}) for any scaling function $\chi(mr)$
describes the behavior of the Green function for $\Lambda r>>1$
and any fixed value of $mr$. The inertial range corresponds
to the additional condition that $mr<<1$. The form of the function
$\chi(mr)$ is not determined by the RG equations themselves; in
the theory of critical phenomena, its behavior for $mr\to0$ is
studied using the well-known Wilson operator product expansion (OPE);
see, e.g., Ref.~\cite{Zinn}. This technique is also applicable to the
theory of turbulence; see, e.g., Ref.~\cite{UFN,turbo}.

According to the OPE, the equal-time product $F_{1}(x)F_{2}(x')$
of two renormalized composite operators at
${\bfx}\equiv ({\bfx} + {\bfx'} )/2 = {\const}$ and
${\bfr}\equiv {\bfx} - {\bfx'}\to 0$ has the representation
\begin{equation}
F_{1}(x)F_{2}(x')=\sum_{a}C_{a} ({\bfr}) F_{a}(t,{\bfx}) ,
\label{OPE}
\end{equation}
where the functions $C_{a}$  are the Wilson coefficients regular
in $m^{2}$ and $F_{a}$ are, in general, all possible renormalized local
composite operators allowed by symmetry; more precisely, the
operators entering into the OPE are those which appear in the
corresponding Taylor expansions, and also all possible operators
that admix to them in renormalization. If these operators have
additional vector indices, they are contracted with the
corresponding indices of the coefficients $C_{a}$.

Without loss of generality it can be assumed that the expansion in
Eq. (\ref{OPE}) is made in basis operators (\ref{2.5}) with
definite critical dimensions $\Delta_{a}$. The renormalized
correlation function $\langle F_{1}(x)F_{2}(x') \rangle$
is obtained by averaging Eq. (\ref{OPE}) with the weight
$\exp S_{R}$, the quantities  $\langle F_{a}\rangle$
appear on the right hand side. Their asymptotic behavior
for $m\to0$ is found from the corresponding RG equations and
has the form $\langle F_{a}\rangle \propto  m^{\Delta_{a}}$.
 From the operator product expansion (\ref{OPE}) we therefore
find the following expression for the scaling function
$\chi(mr)$ in the representation (\ref{RGR}) for the
correlation function $\langle F_{1}(x)F_{2}(x') \rangle$:
\begin{equation}
\chi(mr)=\sum_{a}A_{a}\,(mr)^{\Delta_{a}},
\label{OR}
\end{equation}
where the coefficients $A_{a}=A_{a}(mr)$ are regular
in $(mr)^{2}$; they depend on $\eps$, $d$ and, in our case, on
the cosine $z\equiv\cos\theta=\hat{\bbox{B}}^o\cdot\bbox{r}/r$.

Consider for definiteness the equal-time pair correlation function of
the operators (\ref{Fnp}); their vector indices will be omitted in
order to simplify the notation. For the leading term in the
asymptotic region $\Lambda r>>1$ from the general expression
(\ref{RGR}) we obtain
\begin{equation}
\langle F^{(np)} (x_{1}) F^{(n'p')} (x_{2}) \rangle =
(\Lambda r)^{-\Delta_{n,p_{n}}-\Delta_{n',p_{n'}}}\,
\chi_{np,n'p'}(mr),
\label{pair}
\end{equation}
with the dimensions $\Delta_{n,p}$ from Eq. (\ref{Dnp}) and certain
functions $\chi_{np,n'p'}(mr)$. We recall that the monomial
(\ref{Fnp}) is a linear combination of basis operators possessing
definite critical dimensions (\ref{Dnp}) with different values of
the indices; we also recall that $p_{n}$ is the minimal possible
value of $p$ for a given $n$, i.e., $p_{n}=0$ if $n$ is even and
$p_{n}=1$ if $n$ is odd.
In Eq. (\ref{pair}), only the leading contribution is
displayed which is determined by the {\it minimal} dimensions
entering into the operators on the left hand side (see the
discussion in the end of Sec.~\ref{sec:Operators}).

The leading term of the Taylor expansion for the function
(\ref{pair}) involves the operators $F^{(kl)}$ from (\ref{Fnp}) with
$k=n+n'$ and $l \le p+p'$; higher-order terms involve tensors of
arbitrary rank, built of the field $\B$ and its {\it derivatives}.
The decomposition in renormalized operators gives rise to all the
tensors $F^{(kl)}$ with $k\le n+n'$ and all possible values of $p$;
the tensors with $l > p+p'$ appear owing to the renormalization of
the higher-order terms with derivatives. Therefore, the desired
asymptotic expression for the function $\chi_{np,n'p'}(mr)$
in Eq. (\ref{pair}) in the region $mr<<1$ has the form
\begin{equation}
\chi_{np,n'p'}(mr) = \sum_{k=0}^{n+n'} \sum_{j}
A_{kj} (mr)^{\Delta_{k,j}} + \cdots,
\label{pair2}
\end{equation}
where $A_{kj}$ are coefficients dependent only on $\eps$, $d$ and
$z\equiv\cos\theta$, and the second summation runs over all values
of $j$, allowed for a given $k$.

Some remarks are now in order.

The leading term of the inertial-range behavior ($mr<<1$) of the
function $\chi_{np,n'p'}(mr)$ is obviously given by the
contribution with the minimal dimension $\Delta_{k,j}$ entering
into Eq. (\ref{pair2}).

The dots in Eq. (\ref{pair2}) stand for the contributions of the order
$(mr)^{2+O(\eps)}$ and higher, which arise from the senior operators,
for example, $\B \partial^{2} \B$ and so on.

The operators $F^{(kj)}$ with $k>n+n'$ (whose contributions would be
more important) do not appear in Eq. (\ref{pair2}), because they do
not appear in the Taylor expansion of the function (\ref{pair})
and do not admix in renormalization to the terms of the Taylor
expansion. In other words,
the number of the fields $\B$ in the operator $F_{a}$
entering into the right hand sides of the expansions (\ref{OPE})
can never exceed the total number of the  fields $\B$
in their left hand sides.

The expansion (\ref{pair2}) is consistent with the Legendre polynomial
decomposition of the type (\ref{general}) or, in general, with the
decomposition in irreducible representations of the rotation group,
employed in Refs. \cite{Arad98,Arad99,ABP99}. This becomes especially clear
if the left hand side of Eq. (\ref{OPE}) involves only scalar quantities.
Then all vector indices of the mean values $\langle F_{a}\rangle$ in the
right hand side are contracted with the indices of the corresponding
Wilson coefficients $C_{a}({\bfr})$. As is explained in Sec.
\ref{sec:Operators}, the basis operator that possesses definite
critical dimension $\Delta_{k,j}$ is a $j$-th rank traceless tensor,
so that its mean value is also a $j$-th rank traceless tensor, built
solely of the constant vector $\BO$ and Kronecker delta symbols. It
is then clear that its contraction with $C_{a}({\bfr})$ gives rise
to the $j$-th order Legendre polynomial $P_{j}(z)$.

Now let us turn to the comparison of the
nonperturbative results for the pair correlation function
$C_{\alpha\beta} ({\bfr})=\langle\B_{\alpha}\B_{\beta}\rangle$,
obtained in Sec. \ref{sec:2-point} using the zero-mode techniques,
with the predictions of the RG and OPE, given above. To this end,
we put $n=n'=p=p'=1$ in Eqs. (\ref{pair}), (\ref{pair2}).
The isotropic shell ($j=0$) in Eq. (\ref{pair2}) is then represented
by the trivial operator $F=1$ ($k=0$) with $\Delta_{0,0}=0$ and
the monomial $\B^2\equiv B_{\alpha}B_{\alpha}$ ($k=2$) with
$\Delta_{2,0}=-\eps+O(\eps^{2})$ (see Eqs.~(\ref{Qnp}) and (\ref{Dnp})).
The leading term of the small-$mr$ behavior is given by the latter,
so that we have to identify $\Delta_{2,0}$ with $\zeta_{0}\equiv
\zeta_{0}^{+}$ from Eq. (\ref{expo1}).

It was mentioned in Sec. \ref{sec:Operators} that in the one-loop
approximation, dimensions (\ref{Dnp}) coincide with the critical
dimensions of tensor operators of the type (\ref{Fnp}), but
constructed of the scalar gradients. The above identification
shows that this coincidence is confined to the order $O(\eps)$
even for the simplest dimension $\Delta_{2,0}$. For the scalar
case, one has $\Delta_{2,0}=-\eps$ {\it exactly}, in agreement
with the well-known exact solution for the two-point structure
function obtained in \cite{K68}, while in our case
$\Delta_{2,0}$ is a nontrivial function of $\eps$.

At first sight, the first anisotropic correction is related to the
term with $k=j=1$ in Eq. (\ref{pair2}), i.e., to the simplest
operator $\B$.
However, the mean value $\langle\B(x)\rangle$ vanishes and therefore
gives no contribution to Eq. (\ref{pair2}). Indeed, the analysis of
the diagrams shows that $\langle\B(x)\rangle$ is obtained from the
1-irreducible function $\langle\BP(x)\rangle_{\rm 1-ir}$, which
vanishes owing to the invariance of the model (\ref{action}) with
respect to the shift $\BP\to\BP+\const$.

The leading anisotropic correction is therefore related to the
term with $k=j=2$, i.e., with the operator $B_{\alpha}B_{\beta}$.
Its dimension $\Delta_{2,2}=2\eps/(d-1)(d+2)+O(\eps^{2})$
has to be identified with $\zeta_{2}\equiv \zeta_{2}^{-}$ and is in
agreement with Eq. (\ref{expo2}).

We have thus established the agreement between the $O(\eps)$
results obtained using the RG and OPE, with the first terms
of the expansions in $\eps$ of the exact nonperturbative results
obtained within the zero-mode techniques. Note that for the isotropic
exponent, such agreement was mentioned earlier in Ref. \cite{AA98}
to the order $O(\eps^{2})$.

The exact expressions (\ref{expo1}), (\ref{expo2}) can therefore be
viewed as nonperturbative predictions for the critical dimensions
of the operators $\B^2\equiv B_{\alpha}B_{\alpha}$ and
$B_{\alpha}B_{\beta}$, respectively. Similarly, the results
(\ref{amore}) for the higher exponents $\zeta_{j}^{\pm}$ can be
linked to certain composite operators with two fields $\B$ and
$j$ derivatives for $\zeta_{j}^{+}$ and $(j-2)$ derivatives
for $\zeta_{j}^{-}$. We shall not dwell on this point here and
only note that the exponents $\zeta_{2}^{+}$ and $\zeta_{4}^{-}$
are indeed related to the second-rank and fourth-rank families
of the irreducible operators built of two fields $\B$ and two
derivatives, $\partial B \partial B$, with various arrangements
of the vector indices.

As is explained above in Sec. \ref{sec:2-point}, the exponents
$\zeta_{j}^{\pm}$ for $j\ge4$ do not appear in the inertial-range
behavior of the pair correlation function. This is also easily understood
within the OPE. The mean value of the $j$-th rank irreducible
operator with $n$ fields $\B$ is a traceless $j$-th rank tensor
built of $n$ vectors $\BO$ and Kronecker delta symbols. This
follows from the linearity of the basic equation (\ref{fp}) in $\B$
and $\BO$  (see also the discussion in Sec.~\ref{sec:FT} below
Eqs.~(\ref{lines})). However, nonvanishing tensors of this type
do not exist if the number of vector indices exceeds the number
of fields [the structures like $B^{o}_{\alpha_{1}}B^{o}_{\alpha_{2}}
B^{o}_{\alpha_{3}}B^{o}_{\alpha_{4}}/(B^{o})^{2}$ are forbidden
because $\BO$ appears in the bare propagators (\ref{lines})
only in the numerators].

It was noted in Sec. \ref{sec:2-point} that these exponents will
be activated when a fully anisotropic forcing term (i.e.,
projecting onto all Legendre polynomials) is added to the right
hand side of Eq. (\ref{fp}). Moreover, the above interpretation
in terms of the OPE shows that they are relevant even for the
original simple model (\ref{fp}).

Although the contributions with $j>n$ vanish in the mean value
$\langle B_{\alpha}(x)B_{\beta}(x')\rangle$, they are present in the
expansion (\ref{OPE}) {\it without averaging} and therefore the
exponents $\zeta_{j}^{\pm}$ can reveal themselves in other correlation
functions that involve the product $B_{\alpha}(x)B_{\beta}(x')$. In
particular, they are relevant for the asymptotic behavior of the functions
$\langle B_{\alpha}(x)B_{\beta}(x')\Phi(x_{1})\cdots\Phi(x_{n})\rangle$
for $x\to x'$. Of course, these exponents also appear in the
representations (\ref{RGR}) if the correlation function $F(r)$ in the
left hand side involves the operators with $j>n$.

\section{Discussion and conclusions} \label {sec:Conclusion}

The zero-mode and RG techniques have been exploited in a model
of magnetohydrodinamics turbulence where the magnetic field is
passively advected by a Gaussian white-in-time
velocity, in the presence of a constant background magnetic
field that introduces a large-scale anisotropy.
The basic equations of the model
are Eqs. (\ref{fp})--(\ref{eddydiff}). We have shown that the
correlation functions of the magnetic fluctuations exhibit inertial-range
anomalous scaling.
The explicit asymptotic expressions for the correlation functions of the
magnetic field and their powers have been obtained.
In the inertial range, the correlation functions are represented as
superpositions of power laws with universal exponents and nonuniversal
amplitudes. The anomalous exponents have been calculated
both nonperturbatively (for the second-order correlation function)
and perturbatively (for the second and higher-order correlation functions),
in the first order of the exponent $\eps$ and in any space dimension $d$.

In the language of the zero-mode techniques, anomalous exponents are
associated with scale invariant functions which are annihilated by the
inertial operator ${\cal L}$ (remember that in defining ${\cal L}$ we
neglected the molecular diffusivity $\kappa_0$): the so-called zero
modes of the equations for the correlation functions.
In the language of the RG, these exponents are determined by the
critical dimensions of tensor composite operators
built of the magnetic field without derivatives, Eq. (\ref{Fnp}),
and exhibit a kind of hierarchy related
to the degree of anisotropy: the less is the rank, the less is the
dimension and, consequently, the more important is the contribution
to the inertial-range behavior. The leading terms of the even (odd)
structure functions are given by the scalar (vector) operators.

For the pair correlation function, the complete set of the exponents
has been calculated nonperturbatively using the exact equation
(\ref{eq-moto}); they are given in Eqs.~(\ref{amore}) together with
the discussion of their admissibility.

The general expressions (\ref{RGR}), (\ref{pair})
describe the behavior of the correlation functions for $\Lambda r>>1$,
and any fixed $mr$ ($m\equiv 1/L$)
where $\Lambda^{-1}\propto \eta$, $\eta$ being the dissipative scale,
and $L$ is the integral scale of the problem;
expressions (\ref{OR}), (\ref{pair2})
correspond to the additional condition $mr<<1$ (inertial range).
These results for the leading terms can be summarized as follows:
\begin{equation}
\langle B_{\parallel}^n (t, \bbox{x}) B_{\parallel}^q (t,\bbox{x}') \rangle
\propto (\Lambda r)^{-\Delta_{n,p_{n}}-\Delta_{q,p_{q}}} \,
(mr)^{\Delta_{n+q,p_{n+q}}}
\propto r^{\zeta^{n,q}}, \qquad r = |\bbox{x}-\bbox{x}'|
\label{final}
\end{equation}
with $\Delta_{n,p}$ given by [see Eqs.~(\ref{Qnp}) and (\ref{Dnp})]
\begin{equation}
\Delta_{n,p}
=\eps\,\frac{2p(p-1)-(d-1)(n-p)(d+n+p)}{2(d-1)(d+2)}+O(\eps^{2}).
\end{equation}
Here $B_{||}$ is some component of $\B$, e.g., its
projection onto the direction $\bfr/r$ or $\BO/B^{o}$,
$p_{n}$ is the minimal possible value of $p$ for a given $n$
(i.e.~$p=0$ for $n$ even and $p=1$ for $n$ odd).
The exponents $\zeta^{n,q}$ are expressed through the dimensions
$\Delta_{n,p}$ as follows:
\begin{equation}
\zeta^{n,q} = \cases{ \Delta_{n+q,0}- \Delta_{n,0}- \Delta_{q,0} = -
{\displaystyle \frac{\eps\, nq}{(d+2)}} + O(\eps^{2}) & if $n,q$ are
even, \cr \Delta_{n+q,0}- \Delta_{n,1}- \Delta_{q,1} =  -
{\displaystyle \frac{\eps\, (nq+d+1)}{(d+2)}} + O(\eps^{2}) & if
$n,q$ are odd, \cr \Delta_{n+q,1}- \Delta_{n,0}- \Delta_{q,1} =  -
{\displaystyle \frac{\eps\, nq}{(d+2)}} + O(\eps^{2}) & if $n$
is even and $q$ is odd. \cr}
\label{final1}
\end{equation}

In the presence of an anisotropic forcing, questions about
isotropy restoration at small scales are naturally raised. In
particular, an issue recently addressed concerns the behavior of
the derivative skewness factor of the passive scalar at large P\'eclet
number, $Pe$, in the presence of large-scale anisotropy, and, in a more
general formulation, the effects of large-scale anisotropy on the
inertial-range statistics of passively advected fields
\cite{Sree,synth,A98,CLMV99} and the velocity itself
\cite{PS95,Arad98,Arad99}. In the case of passive advection of a scalar
field, both the real \cite{Sree} and the numerical experiments
\cite{synth,CLMV99} show that the derivative skewness remains $O(1)$ for
very high P\'eclet, in disagreement with what could be expected on the
basis of both dimensional argumentations and cascade ideas.
It means that, contrary to K41 hypothesis, anisotropy
present at large scales persists at small scales.
For the velocity field, in the case of an homogeneous shear flow
an equivalent result has been found for the vorticity, which
keeps a constant value, independent of the Reynolds number \cite{PS95}.

Let us now briefly discuss the consequences of our results for anisotropic
indicators in this problem. Since the equation
(\ref{fp}) is not
invariant with respect to the shift $\B\to\B+\const$, we can
use as the simplest measure of small-scale anisotropy the dimensionless
ratios of the correlation functions of the field $\B$ without
derivatives, e.g.,
\begin{equation}
R_{n} \equiv \frac{\langle B_{||}^{n-1} (\bbox{x}) B_{||} (\bbox{x}')
\rangle} {\langle B_{||} (\bbox{x}) B_{||} (\bbox{x}')\rangle ^{n/2}}.
\label{pair-n}
\end{equation}
 From Eqs.~(\ref{final1}) it then follows that in inertial
range of scales we have:
\begin{mathletters}
\label{skews}
\begin{equation}
R_{2k+1} \propto (\Lambda r)^{-\Delta_{2k,0}} \, (mr)
^{\Delta_{2k+1,1}-(2k+1)\Delta_{2,0}/2}\,,
\label{pair-odd}
\end{equation}
\begin{equation}
R_{2k+2} \propto (\Lambda r)^{-\Delta_{2k+1,1}} \, (mr)
^{\Delta_{2k+2,0}-(k+1)\Delta_{2,0}}\,,
\label{pair-even}
\end{equation}
\end{mathletters}
Note that the ratios (\ref{skews}) depend on both scales of wavenumber
$\Lambda$
and $m$; the dependence on the former follows from the fact that the
powers $B_{||}^{n} $ have nontrivial anomalous dimensions. The dependence
on the P\'eclet number, $Pe \equiv (\Lambda/m)^{\xi}$  can be estimated
by replacing $r$ with $\eta=1/\Lambda$; see Ref.~\cite{Pumir}.
Using explicit $O(\eps)$ expressions for $\Delta_{n,p}$ we then obtain:
\begin{mathletters}
\label{pair-10}
\begin{equation}
R_{2k+1} \propto Pe^{-(d+2-4k^{2})/[2(d+2)]},
\label{pair-11}
\end{equation}
\begin{equation}
R_{2k+2} \propto Pe^{2 k(k+1)/(d+2)}.
\label{pair-12}
\end{equation}
\end{mathletters}

Since the leading terms of the even functions (\ref{final}) are
determined by the exponents of the isotropic shell (i.e., those related
to scalar RG operators), the inertial-range behavior of the even ratios
(\ref{pair-even}), (\ref{pair-12}) is the same as in the isotropic
model.
This gives a quantitative support to the universality of anomalous
exponents
with respect to different classes of forcing.
On the other hand, the odd quantities (\ref{pair-odd}), (\ref{pair-11})
appear to be sensitive to the anisotropy: $R_{3}$ in (\ref{pair-11})
slowly decreases for $Pe\to\infty$, while ratios $R_{2k+1}$ with $k\ge 2$
increase with $Pe$. Moreover, general expressions (\ref{pair-odd})
contain large $\Lambda$ dependent factors, which also prevent these
functions from vanishing at $Pe\to\infty$. Notice the important difference
between the isotropic and the anisotropic problem: in the former
the (nonuniversal) constant of the inertial range power laws of odd-order
moments are zero by symmetry, while this is not the case in the anisotropic
context. This implies that the (hierarchical) exponents for the
odd-order moments appear solely in the anisotropic case.
For a given odd order, the leading exponent is thus responsible
of the observed scale-dependent normalized odd order ratios.

The picture outlined above seems rather general. Indeed it is compatible
with that recently established for the NS
turbulence \cite{Arad98,Arad99} and for the
scalar field passively advected either
by the velocity of the type (\ref{2-point-v}) (see Ref.~\cite{A98})
or by a NS velocity in the two-dimensional inverse cascade
regime (see Ref.~\cite{CLMV99}).
For a passive scalar field advected
by a rapidly changing velocity field such as (\ref{2-point-v}),
RG expressions for
the dimensionless ratios
$R_{n}\equiv S_{n}/S_{2}^{n/2}$, $S_{n}$ being the $n$-th order
structure function of the scalar field, are given by the expressions
(\ref{skews}) {\em without} $\Lambda$ dependent factors and with the same
exponents of $mr$ (see Ref.~\cite{Pumir} for $S_{3}$ and Ref.~\cite{A98}
for
general $n$'s and $d$'s). So, for example, for $k=1$ the ratio
$S_{2k+1}$ decreases down
to the small scales, but much slower than it was expected
on the basis of dimensional argumentations, while for
$k>1$ it grows in agreement with the results of Ref.~\cite{CLMV99}.

It should be emphasized, however, that the results obtained within the
lowest-order approximations in $\eps$ are reliable only for moderate
$n$, because the actual expansion parameter in the Kraichnan model
is $n\eps$ rather than $\eps$ itself (see Ref.~\cite{RG}). The analysis
of the large $n$ behavior requires some additional
resummation of the $\eps$ series, which remains an open problem.
For the passive scalar case, the numerical experiments \cite{CLMV99}
and the instanton calculus \cite{instanton}
show that the exponents analogous to $\zeta^{n,q}$
in Eq. (\ref{final}) tend to a finite limit as $n\to\infty$
(``saturation'').
It is worth noting that the limiting expressions for $n\to\infty$
obtained in Refs. \cite{instanton} diverge as $\xi\to 0$, thus
signalling that small $\xi$ and large $n$ limits do not commute.
The persistence of small-scale anisotropies and the intermittency
saturation
are both statistical signatures of quasi-discontinuities
observed in the scalar field \cite{CLMV99}. It is then reasonable
to expect that the saturation of intermittency
takes place also in the magnetic case where quasi-discontinuous structures
in the magnetic field are likely to be present (see, e.g.,\cite{BW89,PPS89}).

\acknowledgements
It is a pleasure to thank A.~Gruzinov and M.~Vergassola
for their stimulating suggestions on the subject matter.
Useful discussions with L.~Ts.~Adzhemyan, I.~Arad, L.~Biferale,
A.~Celani, R.~Festa, J.-L.~Gilson, J.~Honkonen, I.~Procaccia,
A.~V.~Runov and D.~V.~Vassilevich are also acknowledged.

A.M. was partially supported by the research contract
No.~FMRX-CT-98-0175.
A.L. was supported by the EU contract No.~ERB-FMBI-CT96-0974.
N.V.A. was supported by the Grant Center for Natural Sciences
(Grant No.~97-0-14.1-30) and the Russian Foundation for
Fundamental Research (Grant No.~99-02-16783).

\begin{appendix}
\section{Coefficients in the equations for the magnetic correlation
functions}
\label{blocco1_coeff}
We report hereafter the coefficients $a_i$, $b_i$, $\cdots$, $r_i$
appearing in Eqs.~(\ref{eq1})--(\ref{eq4}).
\subsection{Coefficients in Eq.~(\protect\ref{eq1})}
\begin{eqnarray*}
a_1 &=& d-1\\
b_1 &=& (d-1)(d-1-\xi)\\
c_1 &=& d-1+\xi\\
d_1 &=& (d-1+\xi)(2\xi-d-3)\\
e_1 &=& -\xi^3+3\xi^2+2\xi(d-2)+2d(1-d)\\
f_1 &=& -2d\xi\\
g_1 &=& 2\xi(d-2+\xi)\\
j_1 &=& -\xi[\xi^2+\xi(d-2)-2d]\\
k_1 &=& -2d\xi\\
l_1 &=& -2\xi (2-d-\xi)\\
m_1 &=& -\xi (2\xi^2 -8\xi +8 -2d)\\
n_1 &=& 2\xi\\
o_1 &=& -\xi (\xi -2)\\
p_1 &=& -\xi (\xi -2)(\xi - 4)\\
q_1 &=& \xi (\xi -2)\\
r_1 &=& \xi (\xi -2)(\xi - 4)
\end{eqnarray*}

\subsection{Coefficients in Eq.~(\protect\ref{eq2})}
\begin{eqnarray*}
a_2 &=& (d+\xi-1)[2-2\xi+\xi(\xi-1)]\\
b_2 &=& d-1\\
c_2 &=& (d+\xi-1)(d-1)+2\xi\\
d_2 &=& d+\xi-1\\
e_2 &=& -(d+\xi-1)(d-1)\\
f_2 &=& \xi[\xi^2+\xi(2d-3)+d^2-3d]\\
g_2 &=& 2\xi[\xi^2+\xi(d-2)-d]\\
k_2 &=& d+\xi-1\\
l_2 &=& (d+\xi-1)(\xi-2)\\
m_2 &=& -(d+\xi-1)\\
n_2 &=& -(d+\xi-1)(\xi-2)
\end{eqnarray*}

\subsection{Coefficients in Eq.~(\protect\ref{eq3})}
\begin{eqnarray*}
a_3 &=& (2-\xi)(d+\xi-1)\\
b_3 &=& -\xi(d+\xi-2)\\
c_3 &=& d-1\\
d_3 &=& \xi+(d-1)^2\\
e_3 &=& d+\xi-1\\
f_3 &=& -(d+\xi-1)(d+1)\\
g_3 &=& -d^2+2d-\xi d +4\xi - 1 -2\xi^2\\
j_3 &=& -d \xi\\
k_3 &=& \xi\\
l_3 &=& \xi (d+\xi -2)\\
m_3 &=& -2\xi(\xi-2)\\
n_3 &=& 2\xi(\xi-2)
\end{eqnarray*}

\subsection{Coefficients in Eq.~(\protect\ref{eq4})}
\begin{eqnarray*}
a_4 &=& 2(2-\xi)(d+\xi-1)\\
b_4 &=& d-1\\
c_4 &=& (d+\xi-1)(d-1)+2\xi\\
d_4 &=& d+\xi-1\\
e_4 &=& -(d+\xi-1)(d-1+2\xi)\\
f_4 &=& -2\xi\\
g_4 &=& 2\xi
\end{eqnarray*}

\section{Relations involving the Legendre polynomials}
\label{appa_legendre}

 From the well-known relations involving the Legendre polynomials
$P_j(z)$
(see, e.g., Ref.~\cite{Grad65}) the following decompositions for a
function $F(r,z)=\sum_{j=0}^{\infty}P_j(z) f_{j}(r)$ hold:

\begin{eqnarray}
&&z\partial_z F=\sum_{j=0}^{\infty} P_j
\left[
 j f_j + (2j+1)\sum_{q=1}^{\infty}f_{2q+j}\right]\\
&&z^2\partial_z F= \sum_{j=0}^{\infty}P_j\left[
\frac{j(j-1)}{2j-1}f_{j-1}-\frac{(j+1)(j+2)}{2j+3}
f_{j+1} + (2j+1)\sum_{q=0}^{\infty}f_{2q+j+1} \right]\\
&& z F =\sum_{j=0}^{\infty}P_j\left[
\frac{j}{2j-1}f_{j-1} + \frac{j+1}{2j+3}f_{j+1} \right]\\
&&z^2  F =\sum_{j=0}^{\infty}P_j\left[
\frac{j(j-1)}{(2j-1)(2j-3)}f_{j-2} +
\left(\frac{(j+1)^2}{(2j+3)(2j+1)} + \frac{j^2}{(2j-1)(2j+1)}\right)
f_{j} + \frac{(j+2)(j+1)}{(2j+3)(2j+5)}f_{j+2} \right]\\
&&(1-z^2)\partial_z^2 F = \sum_{j=0}^{\infty}P_j\left[
j(1-j)f_{j} + 2(2j+1) \sum_{q=1}^{\infty}f_{2q+j} \right]\\
&&\partial_z F=  \sum_{j=0}^{\infty}P_j\left[
(2j+1) \sum_{q=0}^{\infty}f_{2q+j+1} \right] .
\end{eqnarray}

\section{Full set of equations projected on the Legendre polynomials}
\label{appa-eq-lunghe}

Inserting Eqs.~(\ref{f125}) and (\ref{f3}) into
Eqs.~(\ref{eq1})-(\ref{eqcont2}) and exploiting the relations
reported in Appendix \ref{appa_legendre}, the following
equations follow from the orthogonality of Legendre polynomials:

\begin{eqnarray}
%
%
&a_1&r^2 f_j^{''(1)}
+b_1 r f_j^{'(1)}  + c_1 \left[j(1-j) f_j^{(1)} +
2(2j+1)\sum_{q=1}^{\infty} f_{2q+j}^{(1)}
\right] + d_1 \left[ j f_j^{(1)} + (2j+1)
\sum_{q=1}^{\infty} f_{2q+j}^{(1)}
\right] +e_1 f_j^{(1)} +\nonumber\\
&f_1& r f_j^{'(2)} +g_1 \left[ j f_j^{(2)} + (2j+1)
\sum_{q=1}^{\infty} f_{2q+j}^{(2)}
\right]+ j_1  f_{j}^{(2)} +
k_1 r \left[\frac{j}{2j-1}f_{j-1}^{'(3)}
+\frac{j+1}{2j+3}f_{j+1}^{'(3)}\right] +
n_1 (2j+1)\sum_{q=0}^{\infty}f_{2q+j+1}^{(3)}+\nonumber\\
&l_1& \left[ \frac{j(j-1)}{2j-1}f_{j-1}^{(3)}-\frac{(j+1)(j+2)}{2j+3}
f_{j+1}^{(3)} + (2j+1)\sum_{q=0}^{\infty}f_{2q+j+1}^{(3)} \right]
+ m_1 \left[ \frac{j}{2j-1}f_{j-1}^{(3)} + \frac{j+1}{2j+3}f_{j+1}^{(3)}
\right] +\nonumber\\
&o_1& f_{j}^{(5)} + p_1 \left[ \frac{j(j-1)}{(2j-1)(2j-3)}f_{j-2}^{(5)} +
\left(\frac{(j+1)^2}{(2j+3)(2j+1)} + \frac{j^2}{(2j-1)(2j+1)}\right)
f_{j}^{(5)} + \frac{(j+2)(j+1)}{(2j+3)(2j+5)}f_{j+2}^{(5)}
\right]= \nonumber\\
&B&^{o\;2}
\left[q_1+r_1 \left(\frac{2}{3}\delta_{j,2}+\frac{1}{3}\delta_{j,0}\right)
\right]\label{eq1proap}\\
\vspace{-1mm}\nonumber\\
%
%
&a_2& f_{j}^{(1)}+ b_2 r^2 f_{j}^{''(2)}
+c_2 r f_{j}^{'(2)}+
d_2 \left[ j(1-j)f_{j}^{(2)} + 2(2j+1)
\sum_{q=1}^{\infty}f_{2q+j}^{(2)} \right]+
e_2 \left[ j f_j^{(2)} + (2j+1)\sum_{q=1}^{\infty}f_{2q+j}^{(2)}\right]
+f_2 f_j^{(2)}+\nonumber\\
&g_2&
\left[ \frac{j}{2j-1}f_{j-1}^{(3)} + \frac{j+1}{2j+3}f_{j+1}^{(3)} \right]+
k_2 f_j^{(5)} +\nonumber\\
&l_2& \left[ \frac{j(j-1)}{(2j-1)(2j-3)}f_{j-2}^{(5)} +
\left(\frac{(j+1)^2}{(2j+3)(2j+1)} + \frac{j^2}{(2j-1)(2j+1)}\right)
f_{j}^{(5)} + \frac{(j+2)(j+1)}{(2j+3)(2j+5)}f_{j+2}^{(5)}
\right]= \nonumber\\
&B&^{o\;2}
\left[m_2+n_2 \left(\frac{2}{3}\delta_{j,2}+\frac{1}{3}\delta_{j,0}\right)
\right]\label{eq2proap} \\
&&\vspace{-1mm}\nonumber\\
&a_3& (2j+1) \sum_{q=0}^{\infty}f_{2q+j+1}^{(1)}+b_3
(2j+1) \sum_{q=0}^{\infty}f_{2q+j+1}^{(2)}+
c_3 r^2 f_{j}^{''(3)}+ d_3 r f_{j}^{'(3)}+
e_3 \left[ j(1-j)f_{j}^{(3)} + 2(2j+1)
\sum_{q=1}^{\infty}f_{2q+j}^{(3)} \right]+ \nonumber\\
&f_3& \left[ j f_j^{(3)} + (2j+1)\sum_{q=1}^{\infty}f_{2q+j}^{(3)}\right]
+ g_3 f_j^{(3)}+j_3 r
\left[ \frac{j}{2j-1}f_{j-1}^{'(5)} + \frac{j+1}{2j+3}f_{j+1}^{'(5)}
\right]
+ k_3 (2j+1) \sum_{q=0}^{\infty}f_{2q+j+1}^{(5)}+\nonumber\\
&l_3&
\left[ \frac{j(j-1)}{2j-1}f_{j-1}^{(5)}-\frac{(j+1)(j+2)}{2j+3}
f_{j+1}^{(5)} + (2j+1)\sum_{q=0}^{\infty}f_{2q+j+1}^{(5)} \right]
+m_3 \left[ \frac{j}{2j-1}f_{j-1}^{(5)} +
\frac{j+1}{2j+3}f_{j+1}^{(5)} \right]
= n_3\,B^{o\;2}\, \delta_{j,2}\label{eq3proap}\\
&&\vspace{-1mm}\nonumber\\
%
%
&a_4& (2j+1) \sum_{q=0}^{\infty}f_{2q+j+1}^{(3)}
+ b_4 r^2 f_{j}^{''(5)}+ c_4 r f_{j}^{'(5)}+
d_4 \left[ j(1-j)f_{j}^{(5)} + 2(2j+1)
\sum_{q=1}^{\infty}f_{2q+j}^{(5)} \right]+\nonumber\\
&e_4& \left[ j f_j^{(5)} + (2j+1)\sum_{q=1}^{\infty}f_{2q+j}^{(5)}\right]
+f_4 f_{j}^{(5)}=g_4\, B^{o\;2} \delta_{j,0}\label{eq4proap}\\
&&\vspace{-1mm}\nonumber\\
%
%
&r&f_{j}^{'(1)} +(d-1)f_{j}^{(1)} + r f_{j}^{'(2)}
-\left[ j f_j^{(2)} + (2j+1)\sum_{q=1}^{\infty}f_{2q+j}^{(2)}\right]
+ r \left[ \frac{j}{2j-1}f_{j-1}^{'(3)}
+ \frac{j+1}{2j+3}f_{j+1}^{'(3)} \right]+\nonumber\\
&&\hspace{-3mm}(2j+1) \sum_{q=0}^{\infty}f_{2q+j+1}^{(3)}-
\left[ \frac{j(j-1)}{2j-1}f_{j-1}^{(3)}-\frac{(j+1)(j+2)}{2j+3}
f_{j+1}^{(3)} + (2j+1)\sum_{q=0}^{\infty}f_{2q+j+1}^{(3)}
\right]-\nonumber\\
&&\hspace{-3mm}\left[ \frac{j}{2j-1}f_{j-1}^{(3)} +
\frac{j+1}{2j+3}f_{j+1}^{(3)} \right]=0\label{eqcont1proap}\\
%
%
&&\vspace{-1mm}\nonumber\\
&&\hspace{-3mm}(2j+1) \sum_{q=0}^{\infty}f_{2q+j+1}^{(2)}+
r f_{j}^{'(3)} + d\, f_{j}^{(3)}+ r
\left[ \frac{j}{2j-1}f_{j-1}^{'(5)} + \frac{j+1}{2j+3}f_{j+1}^{'(5)}
\right]
+ \nonumber\\
&&\hspace{-3mm}(2j+1) \sum_{q=0}^{\infty}f_{2q+j+1}^{(5)} -
\frac{j(j-1)}{2j-1}f_{j-1}^{(5)}+\frac{(j+1)(j+2)}{2j+3}
f_{j+1}^{(5)} - (2j+1)\sum_{q=0}^{\infty}f_{2q+j+1}^{(5)}
= 0. \label{eqcont2proap}
\end{eqnarray}

\end{appendix}

\begin{table}
\caption{Canonical dimensions of the fields and parameters in the
model (\protect\ref{action})}
\label{table1}
\begin{tabular}{cccccccccc}
$F$ & $\B$, $\BO$ & $\BP$ & $ {\bv} $ & $\kappa$, $\kappa_{0}$
& $m$, $\mu$, $\Lambda$ & $D$, $D_{0}$ & $g_{0}$ & $g$ \\
\tableline
$d_{F}^{k}$ & 0 & $d$ & $-1$ & $-2$ & 1& $-2+\eps$ & $\eps $ & 0 \\
$d_{F}^{\omega }$ & 0 & 0 & 1 & 1 & 0 & 1 & 0 & 0 \\
$d_{F}$ & 0 & $d$ & 1 & 0 & 1 & $\eps$ & $\eps$ & 0 \\
\end{tabular}
\end{table}

\begin{references}


\bibitem{EVPP96}
G.~Einaudi, M.~Velli, H.~Politano and A.~Pouquet,
Astrophys. Journ. {\bf 457}, L113 (1996).

\bibitem{K41}
A.~N.~Kolmogorov, CR Acad.~Sci. URSS {\bf 30}, 301 (1941).

\bibitem{Monin} A. S. Monin and A. M. Yaglom, {\it Statistical Fluid
Mechanics} (MIT Press, Cambridge, MA, 1975), Vol.2.

\bibitem{F95}
U.~Frisch, {\em Turbulence. The Legacy of
A.N.~Kolmogorov} (Cambridge University Press, Cambridge, 1995).


\bibitem{Sree} K. R. Sreenivasan, Proc. Roy. Soc. London A
{\bf 434}, 165 (1991); K. R. Sreenivasan and R. A. Antonia,
Annu. Rev. Fluid. Mech. {\bf 29}, 435 (1997).

\bibitem{synth}
M. Holzer and E. D. Siggia, Phys. Fluids {\bf 6}, 1820 (1994);
A. Pumir, Phys. Fluids {\bf 6}, 2118 (1994);
C. Tong and Z. Warhaft, Phys. Fluids {\bf 6}, 2165 (1994).

\bibitem{Sadd94}
S.~G.~Saddoughi and S.~V.~Veeravalli, J. Fluid Mech. {\bf 268}, 333 (1994).

\bibitem{PS95}
A.~Pumir and B.I.~Shraiman, Phys. Rev. Lett. {\bf 75}, 3114 (1995).

\bibitem{Siggia} B. I. Shraiman and E. D. Siggia, Phys. Rev. Lett.
{\bf 77}, 2463 (1996); A. Pumir, B. I. Shraiman and E. D. Siggia,
Phys. Rev. E {\bf  55}, R1263  (1997).

\bibitem{Pumir} A. Pumir, Europhys. Lett. {\bf 34}, 25 (1996);
{\bf 37}, 529 (1997); Phys. Rev. E {\bf  57}, 2914 (1998).

\bibitem{BoOr96}
V.~Borue and S.~A.~Orszag, J. Fluid Mech. {\bf 306}, 293 (1996).

\bibitem{Arad98}
I.~Arad, B.~Dhruva, S.~Kurien, V.~S.~L'vov, I.~Procaccia
and K.~R. Sreenivasan, Phys. Rev. Lett. {\bf 81}, 5330 (1998).

\bibitem{Arad99}
I.~Arad, L.~Biferale, I.~Mazzitelli and I.~Procaccia,
Phys. Rev. Lett. {\bf 82}, 5040 (1999).

\bibitem{A98} N.~V.~Antonov, Phys. Rev. E {\bf 60}, 6691 (1999);
e-Print archive chao-dyn/9907018.

\bibitem{LM99}
A.~Lanotte and A.~Mazzino, Phys.~Rev.~E {\bf 60}, R3483 (1999).

\bibitem{CLMV99} A.~Celani, A.~Lanotte, A.~Mazzino and M.~Vergassola,
e-Print archive chao-dyn/9909038.



\bibitem{Zeldo83}
Ya.~B.~Zeldovich, A.~A.~Ruzmaikin and D.~D.~Sokoloff,
{\em Magnetic fields in Astrophysics} (Gordon \& Breach, New York, 1983).


\bibitem{MHD1} J. D. Fournier, P. L. Sulem and A. Pouquet,
J. Phys. A {\bf 15},  1393 (1982).

\bibitem{MHD2} L. Ts. Adzhemyan, A. N. Vasil'ev and M. Hnatich,
Theor. Math. Phys. {\bf 64}, 777 (1985); {\bf 72}, 940 (1987).

\bibitem{K68}
R.~H.~Kraichnan, Phys. Fluids, {\bf 11}, 945 (1968).

\bibitem{K94}
R.~H.~Kraichnan, Phys. Rev. Lett., {\bf 52}, 1016 (1994).

\bibitem{GK} K. Gaw\c{e}dzki and A. Kupiainen,
Phys. Rev. Lett. {\bf 75}, 3834 (1995);
D.~Bernard, K.~Gaw\c{e}dzki and A.~Kupiainen,
Phys. Rev. E {\bf  54}, 2564  (1996).

\bibitem{Falk1}  M. Chertkov, G. Falkovich, I. Kolokolov and V. Lebedev,
Phys. Rev. E {\bf  52}, 4924 (1995);
M. Chertkov and  G. Falkovich, Phys. Rev. Lett. {\bf 76}, 2706 (1996).


\bibitem{FMV98}
U.~Frisch, A.~Mazzino and M.~Vergassola,
Phys. Rev. Lett. {\bf 80}, 5532 (1998).

\bibitem{RG} L. Ts. Adzhemyan, N. V. Antonov and A. N. Vasil'ev,
Phys. Rev. E {\bf 58}, 1823 (1998); Theor. Math. Phys. {\bf 120},
1074 (1999).

\bibitem{KA68} A.~P.~ Kasantzev,
Zh. {\'E}ksp. Teor. Fiz. {\bf 53}, 1806 (1967)
[Sov. Phys. JETP {\bf {26}}, 1031 (1968)].

\bibitem{V96} M.~Vergassola, Phys. Rev. E {\bf 53}, R\,3021 (1996).

\bibitem{RK97}
I.~Rogachevskii and N.~Kleeorin, Phys. Rev E, {\bf 56}, 417 (1997).

\bibitem{AA98}
L.~Ts.~Adzhemyan and N.~V.~Antonov, Phys. Rev. E {\bf 58}, 7381 (1998).


\bibitem{CFKV}  M.~Chertkov, G.~Falkovich, I.~Kolokolov and
M.~Vergassola, e-Print archive chao-dyn/9906030.

\bibitem{Zinn} J. Zinn-Justin, {\it Quantum Field Theory and
Critical Phenomena} (Clarendon, Oxford, 1989).

\bibitem{UFN} L. Ts. Adzhemyan, N. V. Antonov and A. N. Vasil'ev, Usp.
Fiz. Nauk, {\bf 166}, 1257 (1996) [Phys. Usp. {\bf 39}, 1193 (1996)].

\bibitem{turbo} L. Ts. Adzhemyan, N. V. Antonov and A. N. Vasiliev,
{\it The Field Theoretic Renormalization Group in Fully Developed
Turbulence} (Gordon \& Breach, London, 1999).

\bibitem{Grad65}
I.~S.~Gradshtejn and I.~M.~Ryzhik, {\em Tables of integrals, series
and products} (Academic Press, New York, 1965).

\bibitem{ABP99} I.~Arad, L.~Biferale and I.~Procaccia,
e-Print archive chao-dyn/9909020.

\bibitem{instanton}
M.~Chertkov, Phys. Rev. E {\bf 55}, 2722 (1997);
E.~Balkovsky and V.~Lebedev, Phys. Rev. E {\bf 58}, 5776 (1998).

\bibitem{BW89} D.~Biskamp and H.~Welter,
Phys. Fluids B {\bf 1}, 1964 (1989).

\bibitem{PPS89}
H.~Politano, A.~Pouquet, and P.L.~Sulem,
Phys. Fluids B {\bf 1}, 2330 (1989).

\end{references}
\end{document}